\documentclass[prd,aps,floatfix,nofootinbib,twocolumn,tightenlines,showpacs]{revtex4}
\usepackage{dcolumn}
\usepackage{amsmath}
\usepackage{latexsym}
\usepackage{graphicx}
\usepackage{bm}

\def\co{{\cal O}}

\def\bx{{\bf x}}

\def\svev#1{\left\langle #1\right\rangle}       
\def\tr{{\rm tr}\,}
\def\Tr{{\rm Tr}\,}

\def\Re{{\rm Re\,}}

\long \def \blockcomment #1\endcomment{}

\def\Eq#1{Eq.~(\ref{#1})}
\def\tbeta{\tilde\beta}

\newcommand{\bee}{\begin{equation}}
\newcommand{\ee}{\end{equation}}
\newcommand{\beea}{\begin{eqnarray}}
\newcommand{\eea}{\end{eqnarray}}

\def\ttl#1{``#1''}

\begin{document}

\title{
SU(4) lattice gauge theory with decuplet fermions: Schr\"odinger functional analysis
}
\author{Thomas DeGrand}
\affiliation{Department of Physics,
University of Colorado, Boulder, CO 80309, USA}
\author{Yigal Shamir}
\author{Benjamin Svetitsky}
\affiliation{Raymond and Beverly Sackler School of Physics and Astronomy, Tel~Aviv University, 69978 Tel~Aviv, Israel}

\begin{abstract}
We complete a program of study of SU($N$) gauge theories coupled to two
flavors of fermions in the two-index symmetric representation by performing numerical simulations in SU(4).
The beta function, defined and calculated via the Schr\"odinger functional, runs more slowly than the two-loop perturbative
result. The mass anomalous dimension levels off in strong coupling at a value of about 0.45, rendering this theory unsuitable for walking technicolor.
A large-$N$ comparison of this data with results from SU(2) and SU(3) reveals striking regularities.
\end{abstract}

\maketitle

\section{Introduction}

In the last few years there has been an explosion of interest in
lattice simulations of theories with gauge fields coupled to
a large number of fermionic degrees of freedom, either many flavors of fermions
in the fundamental representation or a few flavors of fermions in
higher-dimensional representations of the gauge group \cite{reviews}.
The primary motivation to study these models is their potential use as
technicolor theories---extensions to electroweak theory with no fundamental
Higgs field~\cite{Hill:2002ap}.
To be a viable candidate, a theory must exhibit spontaneous symmetry breaking,
providing Goldstone bosons to be eaten by the electroweak gauge bosons.
It must also have a slowly running gauge coupling and a mass anomalous
dimension $\gamma_m$ near unity,
in order to give realistic masses to Standard Model fermions while
correctly suppressing flavor-changing neutral currents.

Expectations for candidate theories, based on the two-loop beta function,
are easily stated~\cite{Caswell:1974gg,Banks:1981nn}.
Too many fermionic degrees of freedom render the Gaussian fixed point infrared
stable.
With a smaller number of fermionic fields, asymptotic freedom returns but
the beta function  possesses a zero, signaling the presence of an
infrared-attractive fixed point (IRFP).
Theories with IRFP's are  said to reside within a ``conformal window'' in
the $(N_c,N_f)$ plane, wherein the IR physics displays conformal
invariance and no particle spectrum.
The fermion mass is then
a relevant parameter whose presence induces a mass gap.
Further decreasing the number of fermion fields takes us out of the conformal
window to the QCD-like domain of confinement and chiral symmetry breaking.
Just outside the conformal window,
there may be a borderland where the beta function approaches zero without
actually crossing it; this gives candidates for ``walking technicolor,''
where the running coupling comes to a near-standstill for many decades in
the energy scale, until chiral symmetry breaking eventually sets in.

This paper presents a study of the running gauge coupling and mass anomalous
dimension for SU(4) gauge fields coupled to two flavors of decuplet fermions,
that is, fermions in the two-index symmetric representation.
It is a continuation of our previous work on SU(2) and SU(3) gauge theories
with fermions in the corresponding
representations~\cite{Shamir:2008pb,DeGrand:2008kx,DeGrand:2009hu,DeGrand:2010na,DeGrand:2011qd,DeGrand:2011vp,DeGrand:2012yq}.
Recent interest in these models dates from the proposal by the authors of Refs.~\cite{Sannino:2004qp,Dietrich:2006cm}
that they might make good technicolor candidates.
All three theories possess two-loop IRFP's, raising the possibility that
they might become walking theories when studied nonperturbatively.
In addition, the small number of flavors was seen as favorable from the
point of view of evading precision electroweak constraints.

Using the Schr\"odinger functional (SF)
method~\cite{Luscher:1992an,Luscher:1993gh,Sint:1995ch,Jansen:1998mx,DellaMorte:2004bc,Sint:1998iq,Capitani:1998mq,DellaMorte:2005kg,Lucini:2008vi},
we were able to confirm an IRFP in
the SU(2) theory, placing it within the conformal window~\cite{DeGrand:2011qd}; our most recent result for the
SU(3) theory was the same, but at a lower level of statistical confidence~\cite{DeGrand:2012yq}.
Our calculations of $\gamma_m$ allowed a strong claim, in both cases, that
$\gamma_m$ levels off at strong coupling so that it never exceeds 0.45; thus
neither theory can be used for walking technicolor, regardless of the existence
of the IRFP.
In an effort to find a phenomenologically viable theory with $\gamma_m\simeq1$,
we turn to the SU(4) theory.
Our motivation for this lies in the one-loop expression for $\gamma_m$,
\bee
\gamma_m=\frac{6C_2(R)}{16\pi^2}g^2,
\label{oneloopgamma}
\ee
which is proportional to the quadratic Casimir operator $C_2(R)$ of the fermion
representation; in going from the sextet SU(3) theory to decuplet SU(4),
$C_2(R)$ increases from 10/3 to 9/2, or 35\%.
If the non-perturbative result were to follow this pattern then a value near 1
would be within reach.

As in our earlier work, we apply the SF method to calculate the
nonperturbative beta function.
The original method yields a discrete analogue of the beta
function that gives the change in the running coupling when the length
scale changes by a fixed ratio.
We have noted~\cite{DeGrand:2011qd} that when the coupling runs slowly, as it does in the theories
at hand, an approximate result for the usual beta function can be obtained
directly.
SF techniques also allow one to measure $\gamma_m$ from the volume dependence
of $Z_P$, the renormalization factor of the pseudoscalar density
$\bar\psi\tau^a\gamma_5\psi$.%
\footnote{
  For other applications of the SF method to technicolor candidates,
  see \cite{Appelquist:2007hu,Appelquist:2009ty,Hietanen:2009az,Bursa:2009we,Bursa:2010xn,Hayakawa:2010yn,Karavirta:2011zg,Karavirta:2012qd}.
}

Analyzing these theories via numerical simulations presents a different set of problems
than is seen in QCD~\cite{DeGrand:2009mt}.
The major new feature is the slow running of the coupling.
As we have noted, the slow running simplifies much of the analysis, both of the
beta function and of $\gamma_m$.
On the other hand, it is difficult to tell a slowly running coupling from one that does not run at all,
so that the location of any IRFP may be poorly determined.
The main problem with slow running, however, is that if the bare coupling is
tuned to be strongly interacting at long distance,
then it will be strongly interacting at short distance.
This raises the possibility of strong lattice artifacts,
in particular the appearance of unphysical phase transitions.

Such a transition is present in all our candidate theories.
Our lattice simulations use fermions with a
Wilson-type discretization, which breaks chiral symmetry. Simulations are done
at zero quark mass, but the quark mass is a derived quantity, determined from the axial Ward identity.
Reaching zero fermion mass involves tuning the hopping parameter at fixed bare coupling.
At strong coupling a lattice transition occurs, and when it does,
nowhere does the fermion mass vanish.
Instead, it jumps discontinuously from positive to negative value.
The absence of a massless theory in strong coupling makes it impossible to
apply the SF method there.
To evade this problem, we change the lattice discretization to push the
transition away from the region of bare parameter space
where we wish to run.

Our result for the beta function is similar to what we found in the triplet SU(2)
and sextet SU(3) theories.
The nonperturbative result is consistently smaller in magnitude than the
two-loop estimate, but in this case, as in SU(3),
 we cannot state definitely that it crosses zero.
The result for $\gamma_m$ follows closely the pattern of the other two theories:
It follows the perturbative line, \Eq{oneloopgamma}, for weak coupling but
departs from it and saturates at $\gamma_m\simeq0.45$ (see below for error
estimates).

The outline of the paper is as follows.
In the next section we describe our generalized
lattice action, formulated to suppress lattice artifacts in the
strong-coupling region.
In Sec.~\ref{sec:beta}
we present our results for the beta function and in Sec.~\ref{sec:gamma}
the mass anomalous dimension.
Since we now hold results for the SU($N$) theories with $N=2$, 3, and 4, we
can discuss them in the language of
large-$N$ gauge theories; we show in Sec.~\ref{sec:summary} that the
consistency of the three theories is remarkable.
The appendices present our method of smearing gauge links in SU(4), perturbative
and nonperturbative
tests of our gauge action, and a tabulation of our simulation ensembles.

\section{Generalizing the lattice action\label{sec:action}}

We study the SU(4) gauge theory coupled to two flavors of dynamical fermions in the
symmetric representation (the decuplet) of the color gauge group.
Our techniques are mostly identical to our previous work with the SU(2) and SU(3) theories~\cite{DeGrand:2011qd,DeGrand:2012yq}.
We have already used a generalized gauge action for the SU(3) theory; here we
give a more thorough discussion.

We use the
Wilson fermion action with added clover term~\cite{Sheikholeslami:1985ij}.
The gauge connections in the fermion action are defined with a differentiable hypercubic (nHYP)
smearing \cite{Hasenfratz:2001hp,Hasenfratz:2007rf} of the fundamental links, from which the decuplet
gauge connection for the fermion operator is constructed.%
\footnote{The extension of this smearing
to the gauge group SU(4) is described in Appendix~\ref{sec:hypSU4}.}
We began this project with the usual single-plaquette Wilson gauge action.
The parameters that are inputs to the simulations are
the gauge coupling $\beta=8/g_0^2$
and the fermion hopping parameter $\kappa$, related to the bare mass $m_0$ by
$\kappa=(8+2m_0)^{-1}$. The clover coefficient is set to its tree-level value of unity.

The Schr\"odinger-functional study of the running coupling is carried out
at zero fermion mass, which defines
the critical hopping parameter $\kappa_c(\beta)$.
We define the fermion mass $m_q$ and its related critical hopping parameter $\kappa_c$
 through the axial Ward identity (AWI),
\bee
\partial_t \sum_\bx \svev{A_0^a(\bx,t)\co^a} = 2m_q \sum_\bx \svev{ P^a(\bx,t)\co^a},
\label{eq:AWI}
\ee
where the axial current $A_\mu^a=\bar \psi \gamma_\mu\gamma_5 (\tau^a/2)\psi$, the pseudoscalar density $P^a=\bar \psi \gamma_5 (\tau^a/2)\psi$, and $\co^a$ is a gauge-invariant
wall source at $t=a$.
(See \cite{DeGrand:2010na} for more details.)
With the simplest lattice action---the single-plaquette gauge action and
thin-link Wilson fermions---it turns out to be impossible to set the fermion mass to zero when the gauge coupling is strong.
In that regime there is a first-order phase boundary at which the AWI mass
jumps across zero as $\kappa$ is adjusted, never taking the value zero.
Thus there is no place where the fermions are massless in strong coupling:
The $\kappa_c$ line simply terminates.
This has been observed in theories with many Wilson-type fermions---many fundamental
 flavors~\cite{Iwasaki:1991mr,Iwasaki:2003de,Nagai:2009ip}, or a few flavors of higher-representation fermions~\cite{Catterall:2008qk,Hietanen:2008mr,DeGrand:2010na}.
For our purposes, this phase transition prevents the extension of the
SF calculation to strong coupling.

The transition is a lattice artifact. Its location can be shifted by changing the lattice action.
For triplet SU(2) with $N_f=2$, SF calculations with thin-link fermions
\cite{Hietanen:2009az,Bursa:2009we} were hampered by this transition;
with nHYP clover fermions we pushed the transition back and exposed the IRFP~\cite{DeGrand:2011qd}.
For SU(3) and SU(4) this change of action
turns out to be insufficient and the first order transition remains at fairly weak coupling.

We find that a modification of the {\em gauge\/} action can move
the transition away and allow the study of stronger couplings.
We  supplement the original plaquette term
with an additional plaquette term, constructed with the same link as is used in the fermion
action---a fat link in the higher representation.
The action is
\beea
S_G &=&
\frac{\beta}{8} \sum \Re \Tr U_\mu(x) U_\nu(x+\hat\mu) U_\mu^\dagger(x+\hat\nu) U_\nu^\dagger(x)
\nonumber\\
& &+ \frac{\beta_{10}}{20} \sum \Re \Tr V_\mu(x) V_\nu(x+\hat\mu)
V_\mu^\dagger(x+\hat\nu) V_\nu^\dagger(x), \nonumber \\
\label{eq:soft}
\eea
where $U_\mu(x)$ is the thin link and $V_\mu(x)$ the fat link in the decuplet
representation.
We will refer to this action as a ``soft gauge action.''%
\footnote{We offer a perturbative analysis of soft gauge actions in Appendix~\ref{sec:appPT},
as well as evidence that they do not negate the advantages of fattening the links
in the fermion action.}
We note that the fat plaquette term would appear in a hopping-parameter
expansion in $O(\kappa^4)$; such an induced term has been blamed for the
first-order phase boundary~\cite{Catterall:2008qk}, so it makes sense to try to
adjust its strength.

\begin{figure}[t]
\begin{center}
\includegraphics[width=\columnwidth,clip]{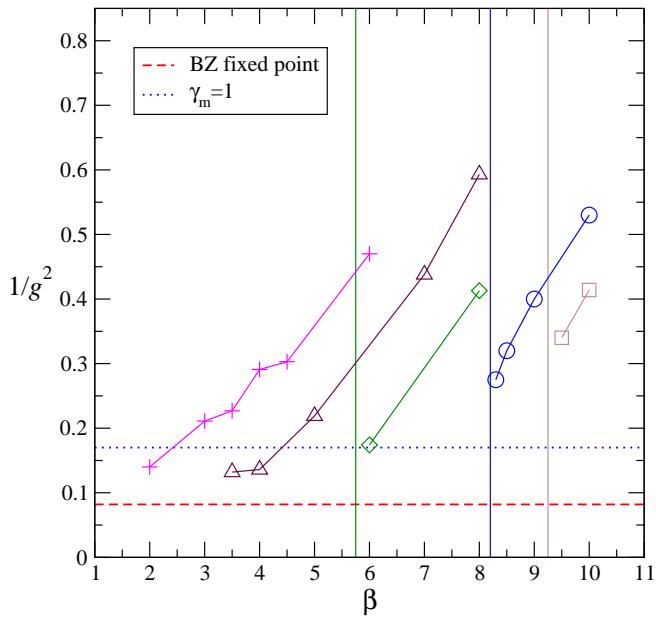}
\end{center}
\caption{Inverse SF coupling versus gauge coupling $\beta$ for several choices
 of $\beta_{10}$, measured with short runs on a $6^4$ lattice.
The connected data sets are for (right to left) $\beta_{10}=-0.5$, 0, 0.5, 1.0, 1.5.
For the first three values of $\beta_{10}$, the vertical lines mark the
appearance of the first-order transition that makes $\kappa_c$ disappear for smaller $\beta$.
The horizontal dashed line near the bottom of the graph marks the
 location of the Banks--Zaks (two-loop) fixed point.
The horizontal dotted line marks where the one-loop $\gamma_m(g^2)$ is equal to unity.
\label{fig:1g2beta}}
\end{figure}

\begin{figure}[t]
\begin{center}
\includegraphics[width=\columnwidth,clip]{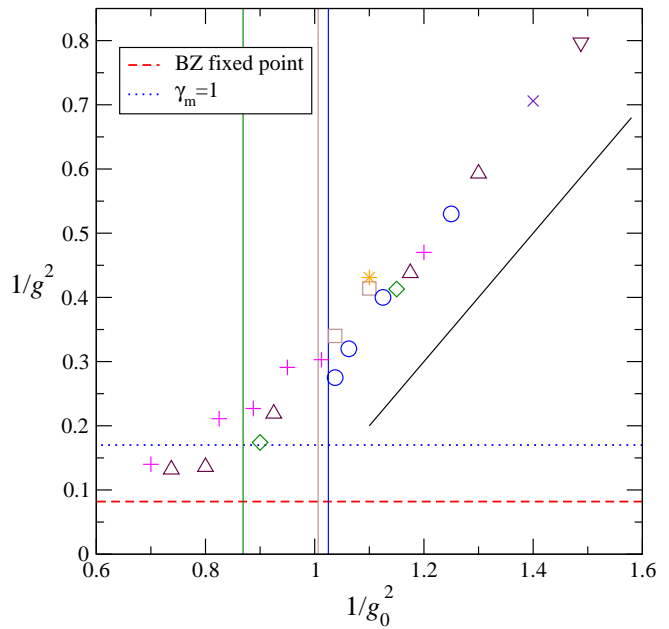}
\end{center}
\caption{Inverse SF coupling versus inverse bare coupling $1/g_0^2$ as defined in
 Eq.~(\protect{\ref{eq:coupling4}}). Plotting symbols are as in Fig.~\protect{\ref{fig:1g2beta}},
plus additional single points for
$\beta_{10}=2.0$ (star),
$\beta_{10}=3.0$ ($\times$),
and
$\beta_{10}=3.5$ (triangle at the top).
The diagonal line has unit slope.
\label{fig:1g21g2s}}
\end{figure}

The advantages of the soft gauge action can be seen in Fig.~\ref{fig:1g2beta},
 a plot of the inverse SF coupling $1/g^2$ versus $\beta$ for various
 values of $\beta_{10}$, measured on a fixed lattice size.
Our goal is to investigate small values of $1/g^2$, in particular the
 neighborhood of the Banks--Zaks fixed point and of the coupling where
 the one-loop anomalous dimension $\gamma_m(g^2)$ is equal to unity.
These are marked by the horizontal lines in the figure.
The circles denote results for $\beta_{10}=0$, meaning the original thin-link plaquette action.
The vertical line just to the left of the leftmost circle marks the bare
 coupling where the first-order boundary appears and there is no $\kappa_c$.
Thus one cannot investigate the region $1/g^2\alt0.25$ with the plaquette action.
Increasing $\beta_{10}$ pushes the phase transition to smaller $\beta$; as it turns out, this
allows us to reach smaller values of $1/g^2$ before encountering the
 transition. For $\beta_{10}\agt1$
we can no longer find the transition in the coupling region we studied.

Having data from many lattice actions allows us to test universality,
 with the results shown in Fig.~\ref{fig:1g21g2s}.
We plot all the lattice data against the perturbative bare lattice coupling,
\begin{equation}
\frac1{g_0^2}=\frac18\left(\beta+\frac{12}5\beta_{10}\right)
\label{eq:coupling4}
\end{equation}
[cf.~\Eq{eq:coupling}].
The points indeed collapse to a common curve, especially in weak coupling.
In perturbation theory, the bare and SF couplings are related by
$g^2= g^2_0 + C g^4_0 + \cdots$, or
\begin{equation}
1/g^2=  1/g^2_0 -C +\cdots.
\end{equation}
The solid diagonal line is plotted to show that the slope of the data indeed
 approaches 1 in weak coupling.

\section{Beta function \label{sec:beta}}

The computation of the running coupling proceeds as described in
Ref.~\cite{DeGrand:2010na}, with adaptations to the SU(4) case.
We set boundary conditions on the gauge fields as described in \cite{Lucini:2008vi},
while the
fermions obey the usual homogeneous boundary conditions at $t=0,L$.
The coupling emerges from a measurement of
the derivative of the action with respect to a parameter $\eta$ in the
boundary gauge field,
\begin{equation}
 \frac{K}{g^2(L)}  =
  \left.\svev{\frac{\partial S_{G}}{\partial\eta}
  -\tr \left( \frac{1}{D_F^\dagger}\;
        \frac{\partial (D_F^\dagger D_F)}{\partial\eta}\;
            \frac{1}{D_F} \right)}\right|_{\eta=0}  .
            \label{deta}
\end{equation}
The constant $K$ can be calculated
from the derivative of the classical continuum action with
respect to $\eta$, giving $K=-12\pi$.
For details of the ensembles generated see Appendix~\ref{sec:ensembles}.

\begin{table}
\caption{Running coupling, \Eq{deta},
evaluated at the bare couplings $(\beta,\beta_{10}, \kappa_c)$ on lattices of size $L$.   The omission of the result for $L=16$ at $\beta=5.0$ is explained
in Appendix~\ref{sec:ensembles}.}
\begin{center}
\begin{ruledtabular}
\begin{tabular}{ccllll}
$\beta$ & $\beta_{10}$ & \multicolumn{4}{c}{$1/g^2$}\\
\cline{3-6}
 &&               $L=6a$    & $L=8a$     & $L=12a$    & $L=16a$  \\
\hline
10.0 & 0 & 0.5778(19) & 0.5440(40) & 0.5241(64) & \hfil--    \\
 9.0 & 0 & 0.4095(19) & 0.3915(37) & 0.3610(42) & 0.3334(72) \\
 8.5 & 0 & 0.3199(19) & 0.2979(32) & 0.2780(69) & \hfil--    \\
 7.0 & 1 & 0.4377(11) & 0.4217(26) & 0.4024(52) & \hfil--    \\
 6.0 & 1 & 0.3101(27) & 0.3019(32) & 0.2844(64) & \hfil--    \\
 5.0 & 1 & 0.2157(27) & 0.2103(40) & 0.2002(53) & \hfil*
\\
 4.5 & 1 & 0.1714(27) & 0.1722(32) & 0.1721(47) & 0.1631(67) \\

 4.0 & 1 & 0.1443(24) & 0.1389(26) & 0.1332(35) & 0.1413(46) \\
 3.5 & 1 & 0.1070(18) & 0.1106(29) & 0.1087(42) & \hfil--    \\
\end{tabular}
\end{ruledtabular}
\end{center}
\label{table:gsf}
\end{table}

\begin{figure}
\begin{center}
\includegraphics[width=\columnwidth,clip]{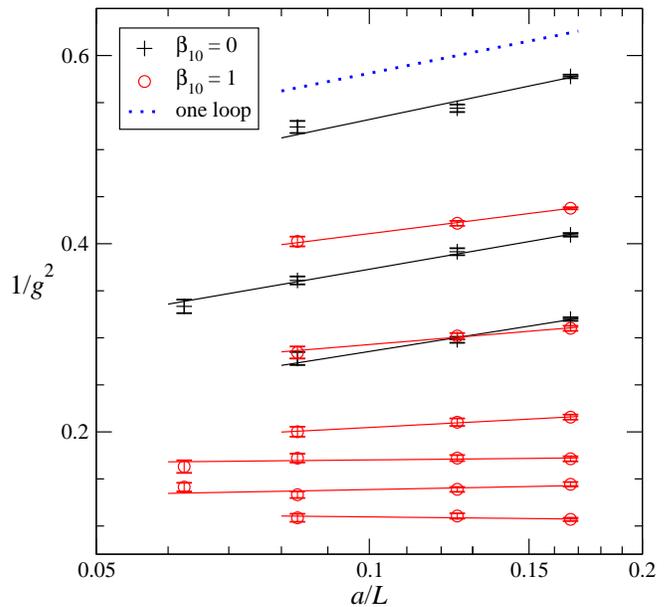}
\end{center}
\caption{
  Running coupling
  $1/g^2$ calculated on lattices of size $L$ (see Table~\ref{table:gsf}).  The crosses are from
  simulations with $\beta_{10}=0$: top to bottom, $\beta=10.0,$ 9.0, 8.5.  The circles are from simulations with
  $\beta_{10}=1$: top to bottom, $\beta=7.0$, 6.0, 5.0, 4.5, 4.0, 3.5.
  The straight lines are linear fits~[\Eq{linfit}]
  to each set of points at given $(\beta,\beta_{10})$;
  the slope gives the beta function.
  The dotted line shows the expected slope from one-loop running.
\label{fig:couplings}}
\end{figure}
Our results for the running coupling are listed in Table~\ref{table:gsf}
and  plotted in Fig.~\ref{fig:couplings}.
It is convenient to define the beta function $\tbeta(u)$ for $u\equiv1/g^2$ as
\bee
  \tbeta(1/g^2) \equiv \frac{d(1/g^2)}{d\log L}
  = \frac{2\beta(g^2)}{g^4} = 2u^2 \beta(1/u)
\label{invbeta}
\ee
in terms of the conventional beta function $\beta(g^2)$.
\begin{figure}
\begin{center}
\includegraphics[width=\columnwidth,clip]{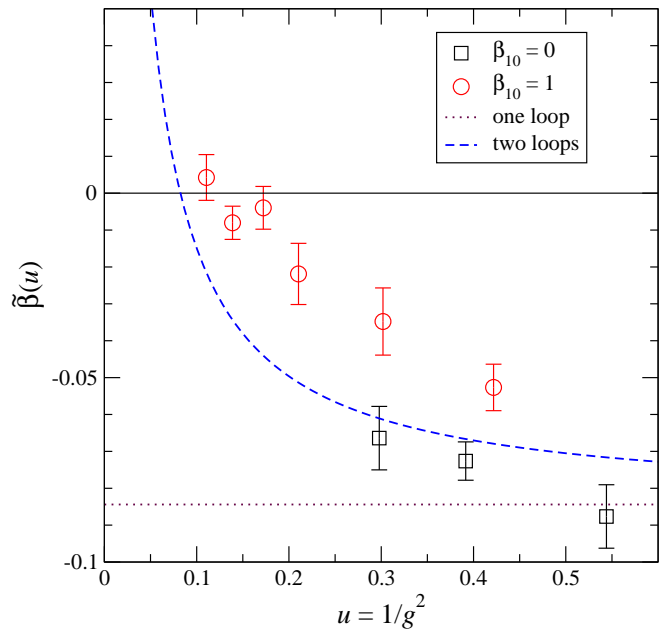}
\end{center}
\caption{Beta function $\tbeta(u)$ as extracted from the
linear fits~(\ref{linfit}), plotted as a function of $u(L=8a)$.
The squares are from the $\beta_{10}=0$ data while the circles
are from $\beta_{10}=1$.
Plotted curves are the one-loop (dotted line)
and two-loop (dashed line) beta functions.
\label{fig:beta}}
\end{figure}
\begin{figure*}
\begin{center}

\includegraphics*[width=.48\textwidth]{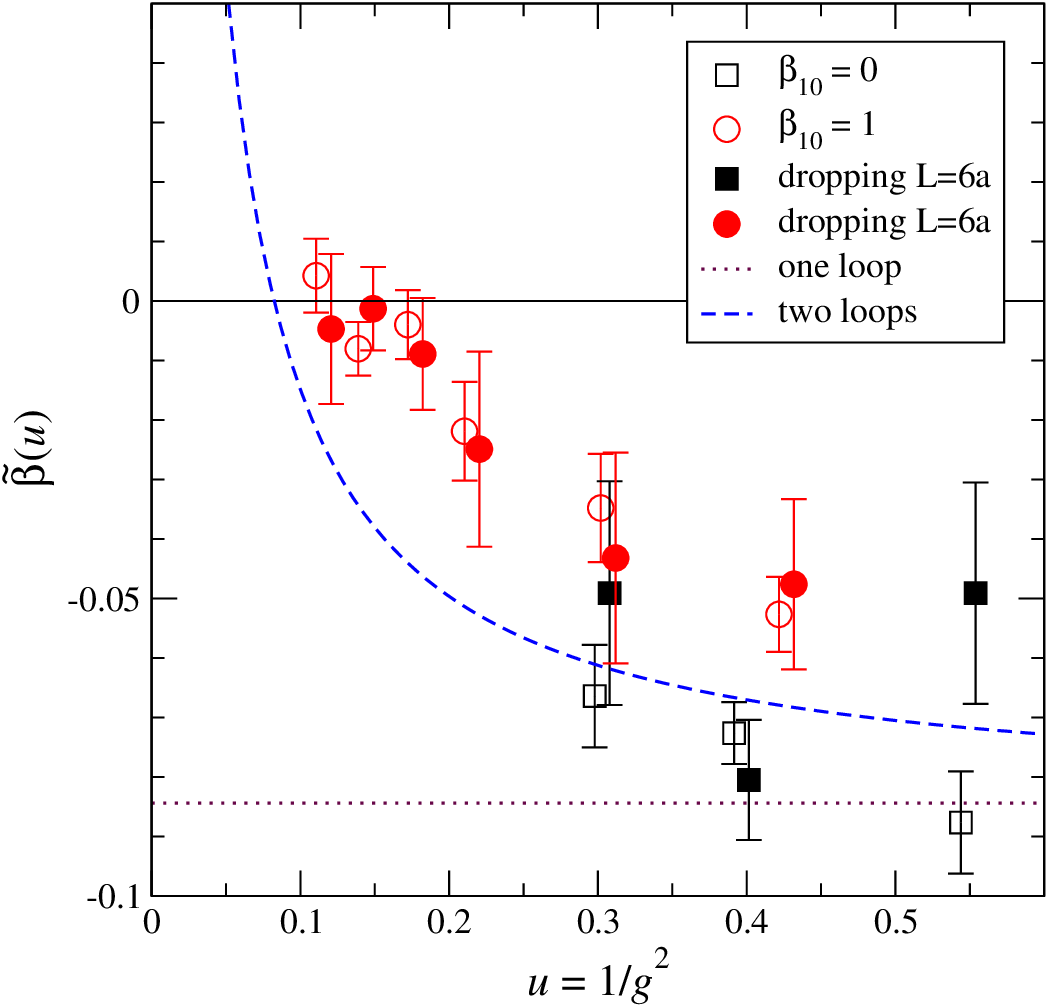}
\hspace{.02\textwidth}
\includegraphics*[width=.48\textwidth]{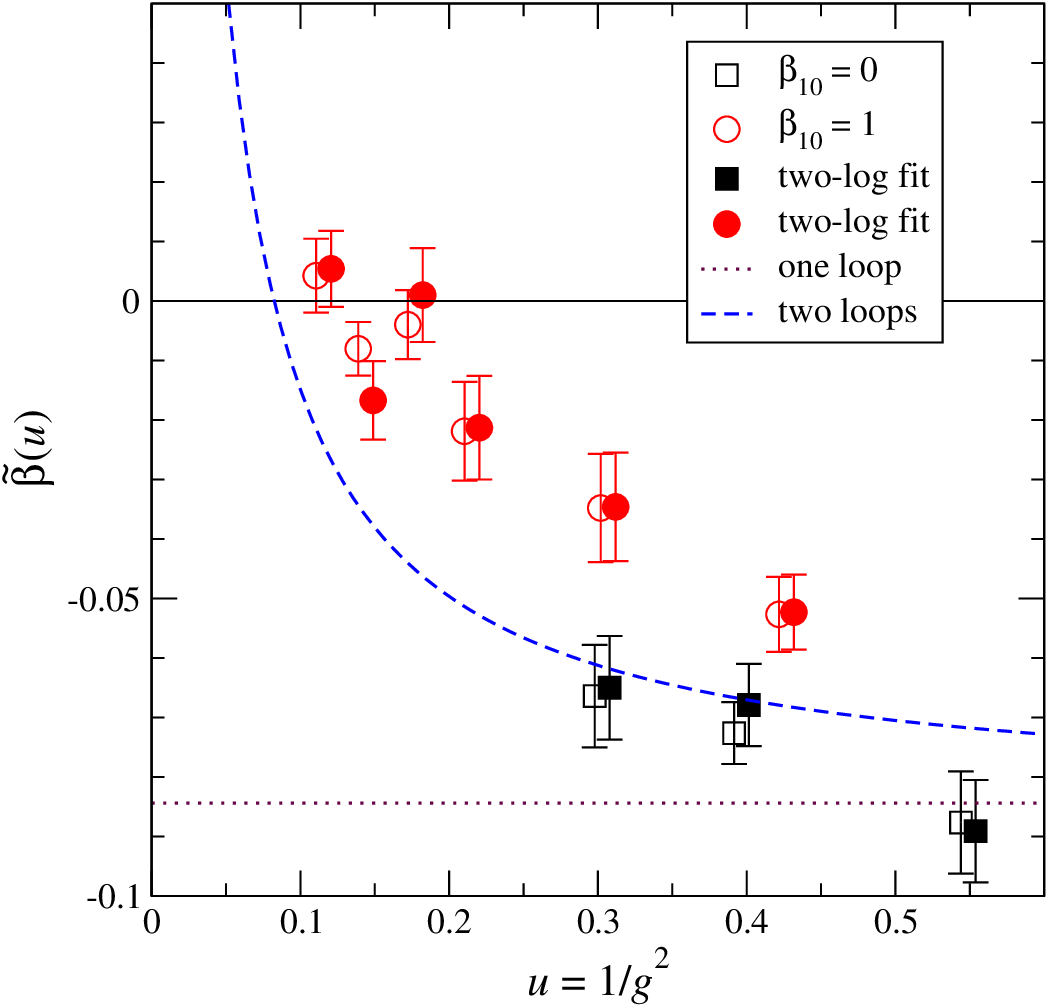}
\vspace*{0ex}
\end{center}
\caption{Comparison of different fit types.
Empty symbols as in Fig.~\ref{fig:beta}.
Left panel: full symbols derive from linear fits in which the $L=6a$ points
are omitted.
Right panel: full symbols derive from fits to \Eq{l2fit},
in which a $\log^2$ term has been added.
Results are plotted against $u(L=8a)$.
The filled symbols have been slightly displaced to the right.
\label{fig:betacompare}}
\end{figure*}
As discussed in Ref.~\cite{DeGrand:2011qd}, the slow running of the
coupling justifies extracting the beta function at each $(\beta,\kappa_c)$ from
a linear fit of the inverse coupling
\bee
u(L/a)=c_0+c_1 \log \frac L{8a} .
\label{linfit}
\ee
$c_1$ is an estimate for the beta function $\tbeta$ at this coupling.
These fits are shown in Fig.~\ref{fig:couplings}.  Each fit was done
using all the available volumes at the given bare parameters.
Values of the beta function $\tbeta(u)$ obtained from these fits
are plotted as a function of $u(L=8a)$ in Fig.~\ref{fig:beta}.
There is some discrepancy between the results for $\beta_{10}=0$
and those for $\beta_{10}=1$; as we shall see, this discrepancy is not
robust under changes of the fitting procedure.
Also shown are the one- and two-loop approximations from the expansion
\begin{equation}
  \tbeta(u) = -\frac{2b_1}{16\pi^2}
  -\frac{2b_2}{(16\pi^2)^2}\frac1u +\cdots ,
\label{btilde}
\end{equation}
where $b_1=20/3$ and $b_2=-260/3$.

The assumption behind the linear fits is that $\tbeta$
is small so that $u(L/a)$ changes very slowly with the volume;
this behavior is apparent in Fig.~\ref{fig:couplings}.
Indeed the fits have good $\chi^2$,
which justifies our hypothesis.
Corrections to the simple model~(\ref{linfit}) come from discretization
errors, as well as from the slight deviation from constancy of the
continuum beta function over the range of volumes.
Discretization errors have the form of powers of $a/L$.
We have found that such corrections are only loosely constrained
in a generalized fit; thus we prefer to estimate the uncertainty due to these
corrections by redoing the linear fits while omitting the smallest lattice,
$L=6a$.
The results are shown in the left-hand panel of Fig.~\ref{fig:betacompare}.
While the error bars have increased, on the whole the results are
stable.%
\footnote{Note that dropping $L=6a$ for the cases where there are only
three volumes leaves no degrees of freedom for the linear fit.}

Deviations from constancy of the (continuum) beta function
give rise to higher powers of $\log L/a$.  Adding the next-to-leading
term, at each bare coupling we fit
\bee
u(L/a)=c_0 +c_1 \log L/8a +c_2 (\log L/8a)^2 .
\label{l2fit}
\ee
From the definition of the beta function it follows that
$c_1$ continues to provide an estimate for the beta function
at $u=1/g^2(L=8a)$.  The results of these fits are shown in the right-hand
panel of Fig.~\ref{fig:betacompare}.  This time there is hardly any change
compared to the linear fits of Eq.~(\ref{linfit}).%
\footnote{Here, too, fitting the cases with only three volumes leaves no degrees
of freedom for the fit.}

While the data trend towards a zero crossing, meaning an IRFP, we
 cannot confirm the existence of this crossing.
It is possible that the beta function turns away from zero, resulting in a walking scenario.
If the function does cross zero, we can offer an estimate of the crossing point.
For each fit type,
we determine the zero of the beta function from a linear fit
of $\tbeta$~vs.~$u$, using the $\beta_{10}=1$ points.
In fitting to any of Figs.~\ref{fig:beta}--\ref{fig:betacompare}, we find little difference whether
we fit to 4, 5, or 6 points starting at the lowest $u$.
Moreover, the three figures give nearly equal central values for the
crossing, and comparable error bars.
The largest $1\sigma$ interval covers all the others.
Taking it as our final uncertainty, we arrive at
\begin{equation}
6.5  \le g_*^2 \le 12
\label{gstar}
\end{equation}
for the location of the supposed IRFP.

\begin{table}
\caption{Pseudoscalar renormalization factor $Z_P$ evaluated at the
  couplings $(\beta,\beta_{10},\kappa_c)$ for lattice
  sizes $L$.}
\begin{center}
\begin{ruledtabular}
\begin{tabular}{ccllll}
$\beta$&$\beta_{10}$
&\multicolumn{4}{c}{$Z_P$}\\
\cline{3-6}
&&   $L=6a$   & $L=8a$     & $L=12a$    & $L=16a$  \\
\hline
10.0 & 0 & 0.2210(3) & 0.2011(6) & 0.1754(7) & \hfil--   \\
 9.0 & 0 & 0.1852(4) & 0.1618(4) & 0.1324(4) & 0.1168(6) \\
 8.5 & 0 & 0.1565(4) & 0.1331(5) & 0.1066(7) & \hfil--   \\
 7.0 & 1 & 0.1869(2) & 0.1682(3) & 0.1473(4) & \hfil--   \\
 6.0 & 1 & 0.1652(5) & 0.14617(35) & 0.1243(7) & \hfil-- \\
 5.0 & 1 & 0.1452(4) & 0.1283(5) & 0.1066(7) & 0.0955(5) \\
 4.5 & 1 & 0.1357(4) & 0.1187(4) & 0.1006(6) & 0.0895(10)\\
 4.0 & 1 & 0.1266(4) & 0.1095(5) & 0.0921(7) & 0.0813(4) \\
 3.5 & 1 & 0.1150(4) & 0.1019(4) & 0.0835(6) & \hfil--   \\
\end{tabular}
\end{ruledtabular}
\end{center}
\label{table:ZP}
\end{table}

\section{Mass anomalous dimension \label{sec:gamma}}

We derive the mass anomalous dimension from the scaling with $L$ of the
pseudoscalar renormalization factor $Z_P$ \cite{DeGrand:2010na,Sint:1998iq,Capitani:1998mq,DellaMorte:2005kg,Bursa:2009we}.
The latter is calculated by taking the ratio
\bee
Z_P = \frac {c \sqrt{f_1}}{f_P(L/2)}.
\label{eq:ZP}
\ee
$f_P$ is the propagator from the a wall source at the $t=0$ boundary
to a point pseudoscalar operator at time $L/2$.
The normalization of the wall source is removed by the $f_1$ factor,
which is a boundary-to-boundary correlator.
The constant $c$, which is an arbitrary normalization,
is $1/\sqrt{2}$ in our convention.

\begin{figure}
\begin{center}
\includegraphics[width=\columnwidth,clip]{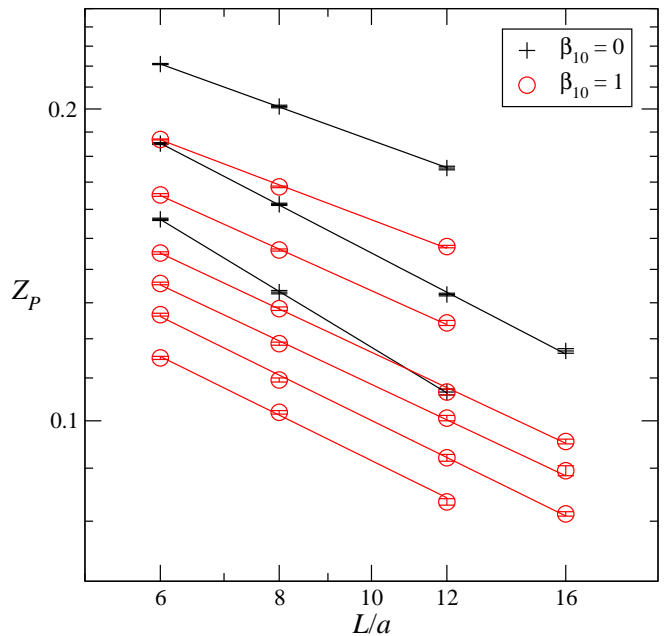}
\end{center}
\caption{The pseudoscalar renormalization constant
  $Z_P$ vs.~$L/a$ (Table~\ref{table:ZP}).
  The crosses are from simulations with $\beta_{10}=0$,
  $\beta=10.0$ to 8.5.  The circles are from simulations with
  $\beta_{10}=1$: top to bottom, $\beta=7.0$ to~3.5.
  The straight lines are fits to each set of points at given
  $(\beta,\beta_{10})$; the slope gives $\gamma_m$.
\label{fig:ZP}}
\end{figure}
\begin{figure}
\begin{center}
\includegraphics[width=\columnwidth,clip]{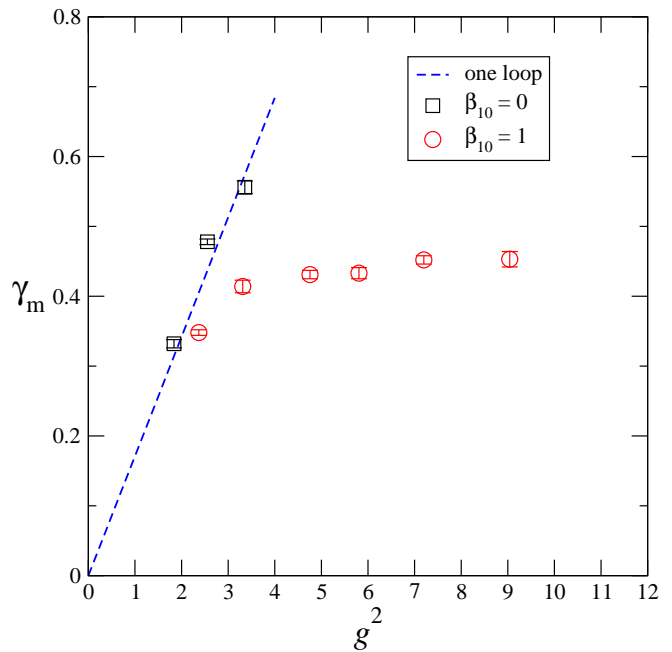}
\end{center}
\caption{Mass anomalous dimension $\gamma(g^2)$ from the linear fits
shown in Fig.~\ref{fig:ZP}, plotted against $g^2(L=8a)$.  The squares
are from the $\beta_{10}=0$ data while the circles are from $\beta_{10}=1$.
The line is the one-loop result.
\label{fig:gamma}}
\end{figure}
\begin{figure*}
\begin{center}

\includegraphics*[width=.48\textwidth]{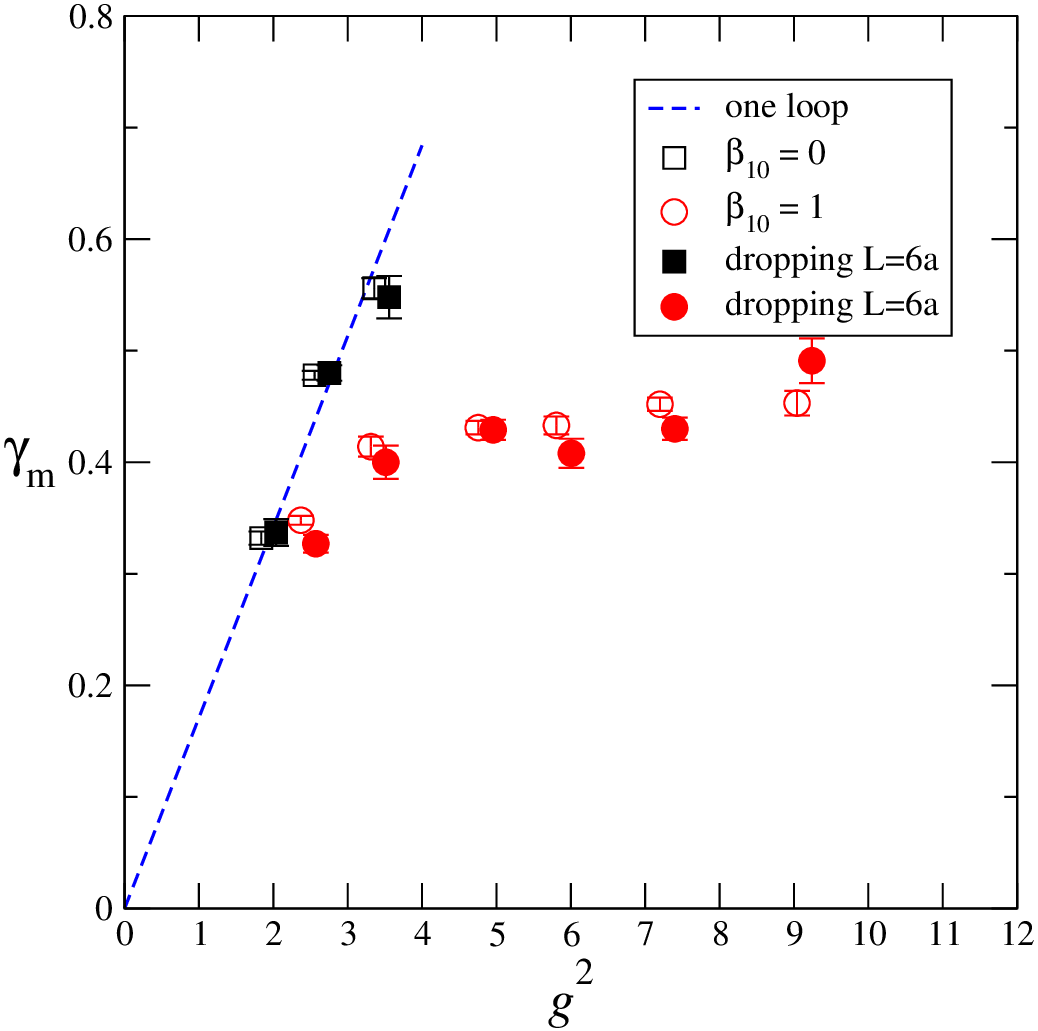}
\hspace{.02\textwidth}
\includegraphics*[width=.48\textwidth]{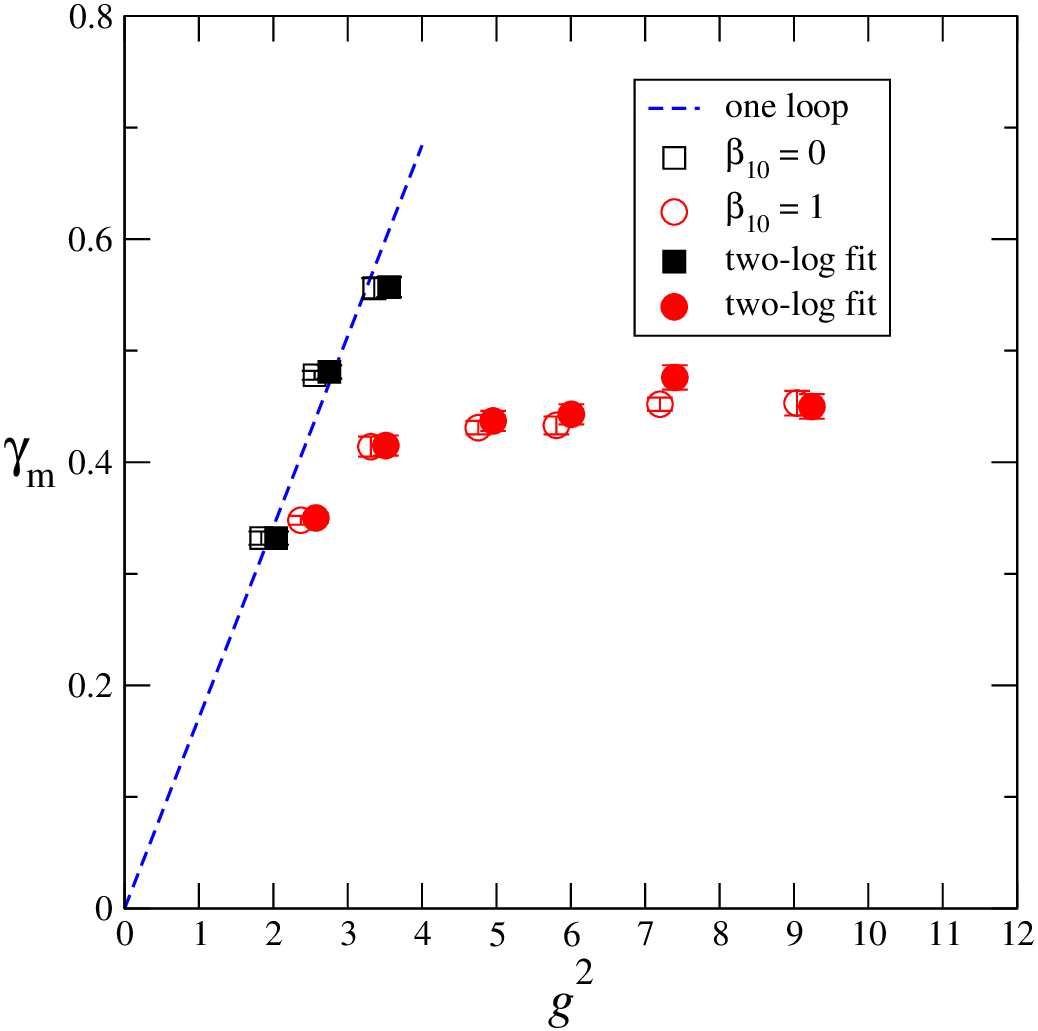}
\vspace*{0ex}
\end{center}
\caption{Comparison of different fit types.
Empty symbols as in Fig.~\ref{fig:gamma}.
Left panel: full symbols derive from linear fits in which the $L=6a$ points
are omitted.
Right panel: full symbols derive from fits to \Eq{l2fitZ},
in which a $\log^2$ term has been added.
Results are plotted against $g^2(L=8a)$.
The filled symbols have been slightly displaced to the right.
\label{fig:gammacompare}}
\end{figure*}

We present the calculated values $Z_P$ in Table~\ref{table:ZP} and plot them in Fig.~\ref{fig:ZP}.
Again \cite{DeGrand:2011qd}, the slow running suggests that we may attempt to extract $\gamma_m$
from the approximate scaling formula
\bee
Z_P(L)=Z_P(L_0)\left(\frac{L_0}L\right)^\gamma ,
\label{Zfp}
\ee
that is, from the slopes of the lines drawn in Fig.~\ref{fig:ZP}.
These linear fits are analogous to Eq.~(\ref{linfit}):
\bee
\log Z_P(L/a)=c_0+c_1 \log \frac L{8a} .
\label{linfitZ}
\ee
The results are shown in Fig.~\ref{fig:gamma}.
Similarly, we have also applied the linear
fit with the smallest volume $L=6a$ removed, and we considered
a fit function analogous to Eq.~(\ref{l2fit}),
\bee
\log Z_P(L/a)=c_0 +c_1 \log L/8a +c_2 (\log L/8a)^2 .
\label{l2fitZ}
\ee
In all cases the mass anomalous dimension
at $g^2(L=8a)$ is given by $-c_1$.
We show a comparison of the different fit types in Fig.~\ref{fig:gammacompare},
plotted against the running coupling $g^2(L=8a)$.
It is apparent that the result for $\gamma_m(g^2)$ is quite robust
under variations in the fitting procedure.
While some of the linear fits to all volumes give high $\chi^2$,
dropping the $L=6a$ points brings $\chi^2$ under control.%
\footnote{There are two $\beta_{10}=1$ points with high $\chi^2$ where dropping $L=6a$ leaves
no degrees of freedom to the fit.
One is the weakest-coupling point; it sits near the one-loop curve
 and agrees with the corresponding $\beta_{10}=0$ point,
so we do not concern ourselves with it further.
The other is the point at the strongest coupling, where $\chi^2/\textrm{dof}=5/1$ before dropping $L=6a$.}

A comparison of $\beta_{10}=0$ to $\beta_{10}=1$ shows that there is some
disagreement.  Of the points obtained with $\beta_{10}=1$,
only the weakest-coupling point is in agreement with the results of
$\beta_{10}=0$ simulations.  The two strongest-coupling points obtained with
$\beta_{10}=0$
lie far above the line connecting the $\beta_{10}=1$ points.
The former originate from simulations near the strong-coupling transition
of the $\beta_{10}=0$ theory.  This is a lattice artifact, pushed off to
stronger couplings by the introduction of $\beta_{10}>0$.
Thus the disagreement between
the two sets of points should be settled in favor of the $\beta_{10}=1$ points.

All our fits show that $\gamma_m$ departs from the one-loop line and levels
off at $\gamma_m\simeq0.45$ at strong coupling.
The highest point in Fig.~\ref{fig:gammacompare} gives us the bound
$\gamma_m<0.51$.

\section{Comparison of SU(2), SU(3),  and SU(4) theories \label{sec:summary}}
The present paper describes one of a set of three related theories, which differ
only in their color content. The obvious way to compare these systems uses the language
of large $N$. The 't~Hooft coupling is $\lambda=g^2N$ and in large $N$ we expect to
see collapse of data for different values of $N$ onto a common function of $\lambda$,
up to $O(1/N)$ corrections.

For theories with $N_f=2$ fermions in the two-index symmetric representation,
the renormalization group equation takes the form
\bee
\frac{d\lambda}{d\log\mu}= -\frac{b_1}{N}\lambda^2 - \frac{b_2}{N^2}\lambda^3+\cdots
\ee
where each of the terms on the right hand side is $O(N^0)$,
\beea
\frac{b_1}{N}&=&\frac{1}{16\pi^2}\left(\frac{7}{3}-\frac{8}{3N}\right),
\label{eq:bfnlowlargeN}\\
\frac{b_2}{N^2}&=&\left(\frac{1}{16\pi^2}\right)^2\left[ \frac23 + O\left(\frac{1}{N}\right) \right].
\eea
The large-$N$ limit of these systems has no IRFP in two-loop order, since $b_2>0$.
The one-loop mass anomalous dimension at $N=\infty$ is
\bee
\gamma_m = \frac{6}{16\pi^2} \lambda.
\label{eq:gmlargeN}
\ee

We collect our data from $N=2$, 3, and 4 and present it in terms of $\lambda$,
taking $g^2$ to be the Schr\"odinger functional coupling.
For clarity, we use only the results of the linear fits, Eqs.~(\ref{linfit})
and~(\ref{linfitZ}), that do {\em not\/} drop $L=6a$.

\begin{figure}
\begin{center}
\includegraphics[width=\columnwidth,clip]{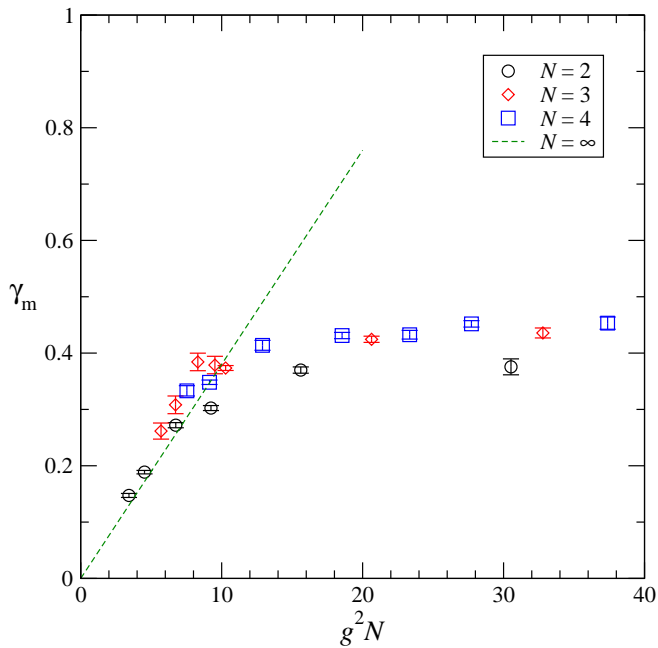}
\end{center}
\caption{ $\gamma_m$ for the SU(2), SU(3), and SU(4) theories displayed as a function of $\lambda=g^2N$ where $g^2$ is the Schr\"odinger functional coupling.
The line is the lowest order large-$N$ prediction, \Eq{eq:gmlargeN}.
For SU(3), we have dropped two strong-coupling points calculated with $\beta_6=0$, since they are
 superseded by $\beta_6=0.5$ data; likewise for SU(4), where two $\beta_{10}=0$ points have been
 dropped (cf.~Fig.~\ref{fig:gamma}).
\label{fig:gammamlargeN}}
\end{figure}

We begin this time with $\gamma_m$, plotted in Fig.~\ref{fig:gammamlargeN}.
We have already noted that all three theories give results for $\gamma_m$
that follow the one-loop line in weak coupling.
Since the rescaled Casimir operators $C_2(R)/N$ for the three theories are
similar (1, 10/9, 9/8) and close to the large-$N$ value [$1+O(1/N)$], the
ascending parts of the curves indeed almost coincide.
What is remarkable is that all three theories fall off the one-loop
line and level off in the neighborhood of $\gamma_m=0.4$.
This is a new regime of large-$N$ scaling behavior.

\begin{figure}
\begin{center}
\includegraphics[width=\columnwidth,clip]{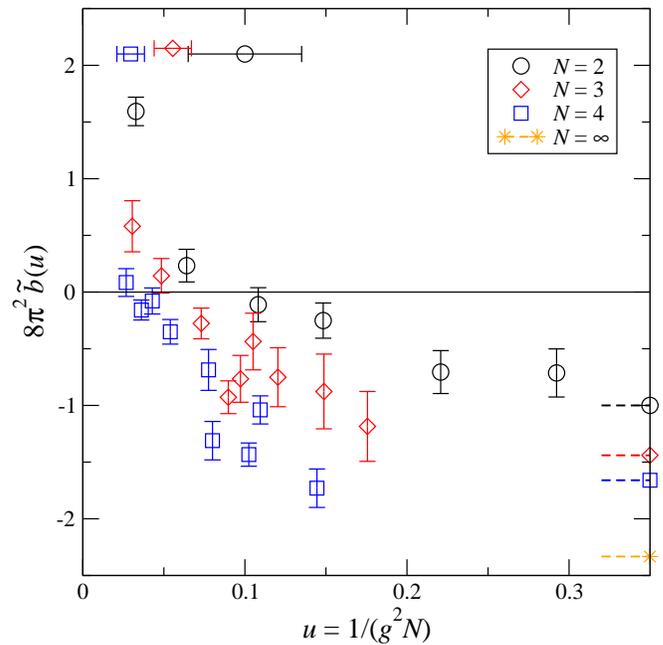}
\end{center}
\caption{Beta functions for the SU(2), SU(3), and SU(4) theories
displayed \`a la large $N$ [\Eq{eq:blargeN}].
The dashed lines at the right edge of the picture mark the weak-coupling limit as predicted by
the one-loop coefficient, \Eq{eq:bfnlowlargeN}.
The points with horizontal error bars mark each beta function's estimated zero  (if it exists).
\label{fig:betalargeN}}
\end{figure}

The beta function, displayed in Fig.~\ref{fig:betalargeN},
shows similar trends.
In analogy with \Eq{invbeta}, we define
\beea
\tilde b(1/\lambda)&=&\frac{d(1/\lambda)}{d\log L}=\frac1N\tbeta(1/g^2)
\label{eq:blargeN}\\
&=&-\frac{2b_1}N-\frac{2b_2}{N^2}\lambda +\cdots,
\eea
and we plot it against $u=1/\lambda$.
For our small values of $N$, the leading
correction to $b_1$ is large, and so the weak coupling limits of the beta
functions for different values of $N$ do not coincide. We have shown the
limiting behavior for each $N$ on the right edge
of Fig.~\ref{fig:betalargeN}, along with the limiting value $-7/(24\pi^2)$ [\Eq{eq:bfnlowlargeN}].
Our results, along with their (possible) fixed points, march leftwards as $N$ increases.
The $N=2$ IRFP lies at $1/\lambda=0.100(35)$. The existence of transitions
for $N=3$ and~4 is more uncertain, but our fits put them at $1/\lambda$ in the range
0.044--0.067 and 0.021--0.038, respectively. It is certainly plausible to imagine
that all three theories have IRFP's, and their location moves to ever
stronger coupling as $N$ increases.
The two-loop beta function predicts that the IRFP has to disappear for $N$ sufficiently large
(at $N\simeq 37$). We cannot rule out that by $N=3$ this has already happened.

\section{Conclusions \label{sec:conclusions}}
We began our study of these related models hoping to answer two questions:
\begin{enumerate}
\item Do the systems exhibit walking, or do they possess an IRFP?
\item How large does the mass anomalous dimension get?
\end{enumerate}
We were able to answer the first question cleanly for $N=2$: there is an IRFP.
For larger $N$, the situation remains less clear: if there is an IRFP, it moves to
ever stronger coupling as $N$ increases, where simulations become ever more difficult.

Qualitatively, the dynamics of all three models are dominated by their slow running.
(This could have been anticipated simply by examining the two-loop beta function.)
This makes the models quite different from conventional QCD with small $N_f$. Slow running was the
key that allowed us to measure the mass anomalous dimension. In all cases, $\gamma_m$
remained less than about 0.45 over the observed range. This observation alone,
independent of the existence of an IRFP, renders the theories unsuitable as candidates
for phenomenologically viable walking technicolor models.

We are struck by the fact that, for all three theories, the two-loop beta function, expressed
in terms of the Schr\"odinger functional coupling, does an excellent job of reproducing
our observed beta functions. At the same time, the behavior of the mass anomalous dimension is quite different
from the perturbative prediction, and is universal---for all three theories $\gamma_m$ falls off
the one loop formula and becomes  (nearly) constant, independent
of the value of the renormalized gauge coupling.

\begin{acknowledgments}
We thank Stefan Schaefer for correspondence, Eli Turkel for advice, and
Evgeny Yurkovsky for assistance.
B.~S. and Y.~S. thank the University of Colorado for hospitality.
This work was supported in part by the Israel Science Foundation
under grant no.~423/09 and by the U.~S. Department of Energy.

The computations for this research were made possible in part by the U.~S.
 National Science Foundation through TeraGrid resources provided by (1) the
 University of Texas and (2) the National Institute for Computational Sciences
(NICS) at the University of Tennessee, under grant number TG-PHY090023.
Additional computations were done on facilities of the USQCD Collaboration at Fermilab,
which are funded by the Office of Science of the U.~S. Department of Energy.
Our computer code is based on the publicly available package of the
 MILC collaboration~\cite{MILC}.
The code for hypercubic smearing was adapted from a program written by A.~Hasenfratz,
R.~Hoffmann and S.~Schaefer~\cite{Hasenfratz:2008ce}.

\end{acknowledgments}
\appendix
\section{$\textbf{n}$HYP smearing for SU(4)\label{sec:hypSU4}}
\subsection{Smearing and normalization}
The formulas of nHYP smearing, introduced in Ref.~\cite{Hasenfratz:2001hp}, are
given in detail in Ref.~\cite{Hasenfratz:2007rf} for the SU(2) and SU(3)
gauge groups.
The smeared links $V_{n\mu}$ are constructed from the bare links $U_{n\mu}$
in three consecutive smearing steps via intermediate fields $\widetilde{V}$
and $\overline{V}$ according to
\begin{widetext}
\begin{subequations}
\begin{eqnarray}
V_{n\mu} & = & \textrm{Proj}_{\textrm{U}(N)}\left[(1-\alpha_{1})U_{n\mu}
+\frac{\alpha_{1}}{6}\sum_{\pm\nu\neq\mu}\widetilde{V}_{n\nu;\mu}\widetilde{V}_{n+\hat{\nu},\mu;\nu}\widetilde{V}_{n+\hat{\mu},\nu;\mu}^{\dagger}\right],\label{eq:HYP-def1}\\
\widetilde{V}_{n\mu;\nu} & = & \textrm{Proj}_{\textrm{U}(N)}\left[(1-\alpha_{2})U_{n\mu}
+\frac{\alpha_{2}}{4}\sum_{\pm\rho\neq\nu,\mu}\overline{V}_{n\rho;\nu\,\mu}\overline{V}_{n+\hat{\rho},\mu;\rho\,\nu}\overline{V}_{n+\hat{\mu},\rho;\nu\,\mu}^{\dagger}\right],\label{eq:HYP-def2}\\
\overline{V}_{n\mu;\nu\,\rho} & = & \textrm{Proj}_{\textrm{U}(N)}\left[(1-\alpha_{3})U_{n\mu}
+\frac{\alpha_{3}}{2}\sum_{\pm\eta\neq\rho,\nu,\mu}U_{n\eta}U_{n+\hat{\eta},\mu}U_{n+\hat{\mu},\eta}^{\dagger}\right].
\end{eqnarray}
\label{eq:HYP-def}
\end{subequations}
\end{widetext}
Here the restricted sums ensure that only links that share a hypercube with
$U_{n\mu}$ enter the smearing.
The parameters $\alpha_i$ offer an arena for optimization but we retained
the choice $(0.75,0.6,0.3)$ of Refs.~\cite{Hasenfratz:2001hp,Hasenfratz:2007rf},
reasoning that the coefficients are geometric in nature and hence shouldn't
change much when the gauge group is changed.
The projection to U($N$) indicated in Eqs.~(\ref{eq:HYP-def}) is of course
dependent on the gauge group.
This is the normalization, the ``n'' in nHYP.

We accomplish the projection by an extension of the method of Ref.~\cite{Hasenfratz:2007rf} to SU(4).
Use of the Cayley--Hamilton theorem gives a formula that can be differentiated
later to obtain the force for the molecular-dynamics evolution.
Given a general $4\times4$ matrix $\Omega$, the projected matrix $V$ is
given by
\begin{equation}
V=\Omega(\Omega^\dagger\Omega)^{-1/2}.
\label{UNproj}
\end{equation}
This requires calculation of the inverse square root of
$Q\equiv\Omega^\dagger\Omega$, which is a positive Hermitian matrix.
Presuming $Q$ to be non-singular, the Cayley--Hamilton theorem allows us
to write $Q^{-1/2}$ as a polynomial in $Q$,
\begin{equation}
Q^{-1/2}=f_0+f_1Q+f_2Q^2+f_3Q^3.
\label{Qhalf}
\end{equation}
We use a Jacobi algorithm to solve for the eigenvalues $g_i$ of $Q$.
(We will not need derivatives of $g_i$ so they need not be found
analytically.)
By writing \Eq{Qhalf} in the eigenbasis of $Q$ we obtain a linear system for
$f_i$,
\begin{equation}
\left(\begin{array}{cccc}
1 & g_{0} & g_{0}^{2} & g_0^3\\[2pt]
1 & g_{1} & g_{1}^{2} & g_1^3\\[2pt]
1 & g_{2} & g_{2}^{2} & g_2^3\\[2pt]
1 & g_{3} & g_{3}^{2} & g_3^3
\end{array}\right)
\left(\begin{array}{c}
f_{0}\\
f_{1}\\
f_{2}\\
f_{3}
\end{array}\right)=
\left(\begin{array}{c}
g_{0}^{-1/2}\\
g_{1}^{-1/2}\\
g_{2}^{-1/2}\\
g_{3}^{-1/2}
\end{array}\right).
\label{eq:matrix-form}
\end{equation}
This system can also be solved numerically; since, however, we will need to
differentiate the result, we solve it analytically by inverting the
Vandermonde matrix.
The solutions $f_i$ are rational
functions of $r_i\equiv \sqrt{g_i}$ or, more
conveniently, of the symmetric polynomials
\begin{subequations}
\begin{eqnarray}
u&=&{r_0}+{r_1}+{r_2}+{r_3}\\[2pt]
v&=&{r_0} {r_1}+{r_0}
   {r_2}+{r_0} {r_3}+{r_1}
   {r_2}+{r_1} {r_3}\nonumber\\&&
  +{r_2} {r_3}\\[2pt]
w&=&{r_0} {r_1} {r_2}+{r_0}
   {r_1} {r_3}+{r_0} {r_2}
   {r_3}+{r_1} {r_2} {r_3}\\[2pt]
x&=&{r_0} {r_1} {r_2} {r_3}.
\end{eqnarray}
\end{subequations}
Denoting a common denominator by
\begin{equation}
\Delta=x \left(u^2 x-u v w+w^2\right),
\end{equation}
we have $f_i=N_i/\Delta$, where
\begin{subequations}
\begin{eqnarray}
N_0&=&u^2 w x-u \left(-v^2 x+v w^2+x^2\right)-v w x\nonumber\\
&&+w^3\\
N_1&=&u^3 x-2 u^2 v w+u \left(v^3+2 w^2\right)\nonumber\\
&&-w \left(v^2+x\right)\\
N_2&=&-u^3 v+u^2 w-u \left(x-2 v^2\right)-2 v w\\
N_3&=&u v-w.
\end{eqnarray}
\end{subequations}
This completes the calculation of the quantities needed to normalize $\Omega$.

\subsection{Force in molecular dynamics}
We follow still the derivation in Sec.~3 of Ref.~\cite{Hasenfratz:2007rf},
which is based on Morningstar and Peardon~\cite{Morningstar:2003gk}.

The force is the derivative of the effective action with respect to
simulation time $\tau$.
The first step is to note that the fermionic part of the action includes
only the fat links $V_{n\mu}$, so
\begin{equation}
\frac d{d\tau}S_{\textrm{eff}}=
\Re\tr\frac{\delta S_{\textrm{eff}}}
{\delta V_{\mu}}\frac{{dV}_{\mu}}{d\tau}
\equiv\Re\tr(\Sigma_{n\mu}\dot{V}_{n\mu}).\label{eq:force1}
\end{equation}
One proceeds to apply the chain rule repeatedly to $\dot{V}_{n\mu}$ via
Eqs.~(\ref{eq:HYP-def}) until one arrives at derivatives $\dot{U}_{n\mu}$ of the
thin links.%
\footnote{In fact the chain rule is first applied to the change of representation from decuplet to fundamental; this is followed by the fat-link chain rule.}
The only factor in the chain rule that depends on the group comes from the
U($N$) projection (\ref{UNproj}), which appears at every level of smearing
in Eqs.~(\ref{eq:HYP-def}).
In order to write a derivative $\dot{V}$ in terms of $\dot{\Omega}$, we
use the Cayley--Hamilton formula (\ref{Qhalf}) [cf.~Eq.~(3.10) of
Ref.~\cite{Hasenfratz:2007rf}],
\begin{eqnarray}
\Re\tr\Sigma\dot{V} & = &
\Re\tr\Bigl[\Sigma\frac d{d\tau}(\Omega Q^{-1/2})\Bigr]\nonumber \\
&=& \Re\tr(Q^{-1/2}\Sigma\dot{\Omega})
+\tr(\Sigma\Omega)\dot{f}_{0}
+\tr(Q\Sigma\Omega)\dot{f}_{1}\nonumber\\&&
+\tr(Q^{2}\Sigma\Omega)\dot{f}_{2}
+\tr(Q^3\Sigma\Omega)\dot f_3+f_{1}\,\tr(\Sigma\Omega\dot{Q}) \nonumber\\
&& 
+f_{2}\,\tr[(\Sigma\Omega Q+Q\Sigma\Omega)\dot{Q}]
+f_3\,\tr[(\Sigma\Omega Q^2 \nonumber\\
&& +Q\Sigma\Omega Q+Q^2\Sigma\Omega)\dot Q].
\nonumber\\
\label{eq:force4}
\end{eqnarray}
Upon defining the traces
\begin{equation}
c_n=\frac1{n+1}\tr Q^{n+1},
\end{equation}
one can write
\begin{equation}
\dot{f}_{i}=\sum_{n=0}^3 b_{in} \tr\big(Q^{n}\dot{Q}\big).
\end{equation}
where
$b_{in}=\partial f_{i}/\partial c_{n}$.
Then
\begin{equation}
\Re\tr\Sigma\dot{V} =
\Re\tr(Q^{-1/2}\Sigma\dot{\Omega})
+\Re\tr A\dot{Q},
\label{eq:force5}
\end{equation}
where
\begin{eqnarray}
A&=&\sum_{n=0}^3\tr(B_{n}\Sigma\Omega)Q^{n}
+f_1\Sigma\Omega+f_2(\Sigma\Omega Q+Q\Sigma\Omega)\nonumber\\
&&+f_3(\Sigma\Omega Q^2+Q\Sigma\Omega Q+Q^2\Sigma\Omega),
\end{eqnarray}
with $B_n=b_{0n}+b_{1n} Q+b_{2n} Q^2+b_{3n} Q^3$.
Now we differentiate $Q=\Omega^\dagger\Omega$ to obtain finally
\begin{equation}
\Re\tr\Sigma\dot{V} =
\Re\tr\left[ (Q^{-1/2}\Sigma+A\Omega^{\dag}+A^{\dag}\Omega^{\dag})\dot{\Omega}\right].
\end{equation}

The derivatives $b_{ij}$ are calculated via the eigenvalues $g_k$
through the chain rule,
\begin{equation}
b_{ij}=\frac{\partial f_i}{\partial c_j}
=\sum_k \frac{\partial f_i}{\partial g_k}\frac{\partial g_k}{\partial c_j}.
\end{equation}
The matrix ${\partial g_k}/{\partial c_j}$ is the inverse of the Vandermonde
matrix ${\partial c_k}/{\partial g_j}=(g_j)^k$, so that we still don't need an
analytical expression for the eigenvalues.
The derivatives ${\partial f_i}/{\partial g_k}$ can be calculated directly from
the above expressions for $f_i$.
The final result for $b_{ij}$ can, like the $f_i$, be written as rational
functions of the symmetric polynomials $u,v,w,x$.
The expressions are lengthy and not particularly illuminating~\cite{ancillary}.

\section{Tests of soft gauge actions\label{sec:appPT}}

In lowest order, the action of \Eq{eq:soft}  is just a quadratic form in the
vector potentials of the thin and fat links.
In momentum space the fat link's gauge field $B_\mu(q)$ is related to the
thin link's gauge field $A_\mu(q)$ through a form factor $\tilde h_{\mu\nu}(q)$
whose specific form depends on the particular definition of the fat link,
\bee
\label{eq:momspA}
 B_\mu(q) = \sum_{\nu} \tilde h_{\mu\nu}(q) A_\nu(q).
\ee
In general,
$\tilde h_{\mu\nu}(q) \sim 1 + O(a^2 q^2) + O(a^4 q^4)+\cdots$.

Let us generalize to $N$ colors, with an ordinary plaquette term made of thin link
 plus a closed loop made of fat
links in representation $R$ of the gauge group.
We can write the quadratic gauge action as
\beea
S_0^G &=& -\frac{1}{2 g_0^2 }\int_{p p^\prime}   (2\pi)^4\delta^4(p+p^\prime)
  \nonumber\\
&& \times  \left[ (1-s) A_\mu(p^\prime)D^{\textrm{thin}}_{\mu\nu}(p) A_\nu(p)
\right.\nonumber\\
&&\left.+s B_\mu(p^\prime)D^{\textrm{fat}}_{\mu\nu}(p) B_\nu(p)\right],
\eea
where $D^\textrm{thin}$ is the thin link kernel and
 $D^\textrm{fat}$ is the fat link kernel.
(We have suppressed color indices.) In practice, we will use a plaquette fat link
term, so $D^{\textrm{fat}} = D^{\textrm{thin}}$.
The bare coupling is
\bee
\frac{1}{g_0^2} = \frac{\beta}{N}T(f)  + \frac{\beta_R}{d_R}T(R),
\label{eq:coupling}
\ee
where $T(f)=1/2$ is the trace normalization of the fundamental representation
while $T(R)$ is that of the representation $R$; $d(R)$ is the dimension of the latter.
We also define $s=x/(1+x)$  where
\bee
x=\frac{\beta_RT(R)/d_R}{\beta T(f)/N}.
\ee
In SU(4) with decuplet fat links,  $x=\frac{12}{5} \beta_R/\beta$.

Upon using \Eq{eq:momspA}, we find the free gauge boson action to be
\bee
S_0^G = -\frac{1}{2 g_0^2 }\int_{p p^\prime}    (2\pi)^4\delta^4(p+p^\prime)
   \left[  A_\mu(p^\prime)D_{\mu\nu}(p) A_\nu(p)\right],
\ee
where
\bee
D_{\mu\nu} = (1-s)D^{\textrm{thin}}_{\mu\nu}
 + s \tilde h_{\rho\mu}D^{\textrm{fat}}_{\rho\sigma}\tilde h_{\sigma\nu}\,.
\ee
After adding a gauge fixing term,
\bee
S_{\rm gf} = -\frac{1}{2g_0^2} \sum_{\mu\nu}\int_k
 \Tr \frac{1}{\xi} \hat k_\mu \hat k_\nu  A_\mu(-k)A_\nu(k),
\ee
where $\hat k_\mu = (2/a) \sin(ak_\mu/2)$,
the gauge boson propagator is found by inverting the field equation,
\bee
\sum_\nu \left[ \frac{1}{\xi} \hat k_\mu \hat k_\nu + D_{\mu\nu}(k)\right ]
   G_{\nu\tau}(k) = \delta_{\mu\tau}  .
\label{eq:gprop}
\ee

The case of a Wilson action and APE smearing can be treated analytically.
In that case both the action and smearing are built out of projectors,
\bee
P^T_{\mu\nu} = \delta_{\mu\nu} - \frac{\hat k_\mu \hat k_\nu}{\hat k^2}
,\qquad
P^L_{\mu\nu} =  \frac{\hat k_\mu \hat k_\nu}{\hat k^2} .
\ee
The Wilson gauge action is
\bee
 D_{\mu\nu}(k) = \hat k^2 P^T_{\mu\nu},
\ee
and the APE smearing term is \cite{Bernard:1999kc}
\bee
\tilde h_{\mu\nu}(q) = f(q)P^T_{\mu\nu} + P^L_{\mu\nu},
\ee
with $f(q)=1-(\alpha/6)\hat q^2$; $\alpha$ is the smearing parameter.
This means that the soft APE-smeared Wilson action is
\bee
 D_{\mu\nu}(k) = (1-s+sf^2) \hat k^2 P^T_{\mu\nu}
\ee
and thus the propagator is
\bee
 G_{\mu\nu}(k) = \frac{1}{\hat k^2}\left[ \frac{1}{1-s+sf^2} P^T_{\mu\nu} + \xi P^L_{\mu\nu}\right] .
\label{eq:wprop}
\ee
More complicated actions and smearing do not have this projector form, and the
final result for $ G_{\mu\nu}(k)$ is  not illuminating.

So much for the gluon propagator. The second ingredient needed for
 perturbative calculations is the
  fermion--gauge boson vertex. In a fat action with unitary links it is
simply the smearing form factor multiplied by the unsmeared vertex $\Gamma_\nu$,
\bee
\tilde \Gamma_\mu = \Gamma_\nu \tilde h_{\nu\mu}.
\ee
All Feynman graphs with an internal gluon line are built of the combination of terms
\bee
\tilde G_{\mu\nu}(k) = \tilde \Gamma_\mu G_{\mu\nu}(k) \tilde \Gamma_\nu
= \Gamma_\nu[   \tilde h_{\nu\mu}  G_{\mu\rho} \tilde h_{\sigma\rho}  ]\Gamma_\sigma
\ee
and so for all practical purposes, the gluon propagator $G_{\mu\nu}(k)$ is just replaced
by the combination $\tilde h_{\nu\mu}  G_{\mu\rho} \tilde h_{\sigma\rho}$.
For APE smearing on top of the Wilson gauge action,
 this replacement can be performed analytically, and the transverse
part of the propagator times form factors is just the usual Wilson  propagator, rescaled by
the factor
\bee
\frac{f^2(q)}{1-s+sf^2(q)}.
\label{eq:fullprop}
\ee

Let us recall some features of perturbation theory with fat links.
Consider first the usual smeared fermion action with unsmeared gauge action, $s=0$.
The good features of this action arise
 because the form factor in \Eq{eq:fullprop} suppresses the large-$q$ part of loop integrals.
This shrinks the size of tadpole graphs and lowers the size of one loop matching factors.
A positive weight $s$ increases the form factor at large $q$, and a negative $s$
 effectively amounts to increased smearing.

The case $s=1$ is peculiar in that the effects of smearing are completely undone.
What is happening in this case is that the transformation $A\rightarrow B$
is just a field redefinition: we are doing Monte Carlo with the fat link
everywhere as the fundamental gauge field.
Only if the gauge action is smeared differently from the vertex can smearing
achieve its goal.

One way to reduce the value of the effective gauge boson propagator at large $q$
is to make $s$ negative. Then the fat term amounts to a further
smearing of the effective quark--gluon vertex. This might be a
feature worth exploring in
low-$N_f$ situations.

As described in the main body of the paper, we
have chosen $s>0$ for a nonperturbative reason, and so we should check if this choice
is benign from the point of view
of perturbation theory.
We show the results for two observables to check this point.

\begin{figure}
\begin{center}
\includegraphics[width=\columnwidth,clip]{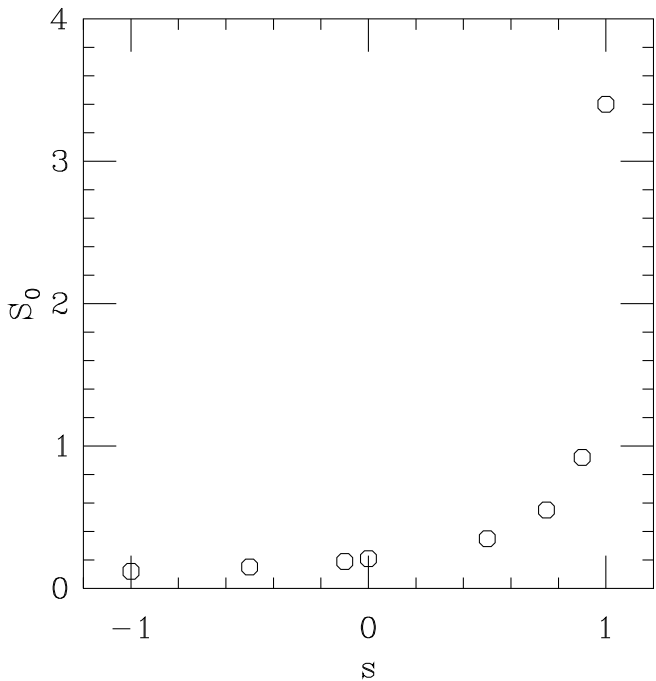}
\end{center}
\caption{Additive mass renormalization
[see \Eq{eq:deltam}] for nHYP clover fermions with $c_{\textrm{SW}}=1$
as a function of the relative weight $s$ of thin- and fat- link
terms.
\label{fig:s0}}
\end{figure}

Fig.~\ref{fig:s0} shows the additive mass renormalization for nHYP clover fermions with $c_{\textrm{SW}}=1$,
parametrized as
\bee
\delta m = \frac{g_0^2}{4\pi} 2T(R) S_0.
\label{eq:deltam}
\ee
(The choice of factors is taken to make the case of fundamental representation fermions most
transparent.)
The $s=1$ case reduces to the thin link value. The undoing of smearing is quite abrupt,
and nearly any value  away from $s=1$ produces a large suppression of $\delta m$.

\begin{figure}
\begin{center}
\includegraphics[width=\columnwidth,clip]{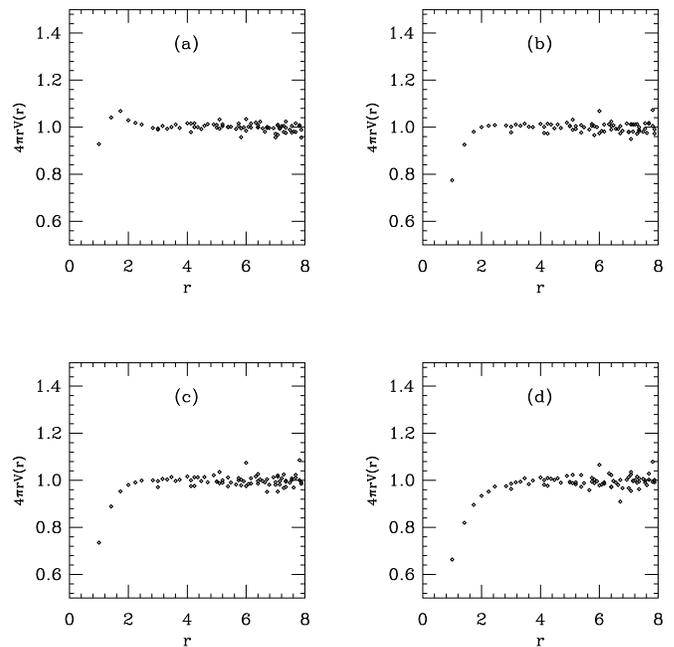}
\end{center}
\caption{$4\pi r V(r)$, the tree-level static potential due to a soft Wilson action and
seen by nHYP-smeared sources, at
 four $s$ values: (a) $s=0.7$, (b) $s=0.2$, (c) $s=0$, (d) $s=-0.5$.
\label{fig:vr}}
\end{figure}

In perturbation theory, the tree level static potential between point sources is Coulombic,
up to lattice artifacts:
\bee
\hat V(r)= g^2 C_F V(r) \sim g^2 C_F\frac{1}{4\pi r}\,.
\ee
When the sources are themselves nHYP-smeared, the short distance part of the potential is softened.
This is a usual consequence of smearing~\cite{Hasenfratz:2001tw}.
Fig.~\ref{fig:vr} shows $4\pi r V(r)$ for soft actions with several choices of $s$.
It appears that from the point of view of perturbation theory, the soft
 gauge action does not obviously harm useful properties of the lattice action.

\begin{figure}
\begin{center}
\includegraphics[width=\columnwidth,clip]{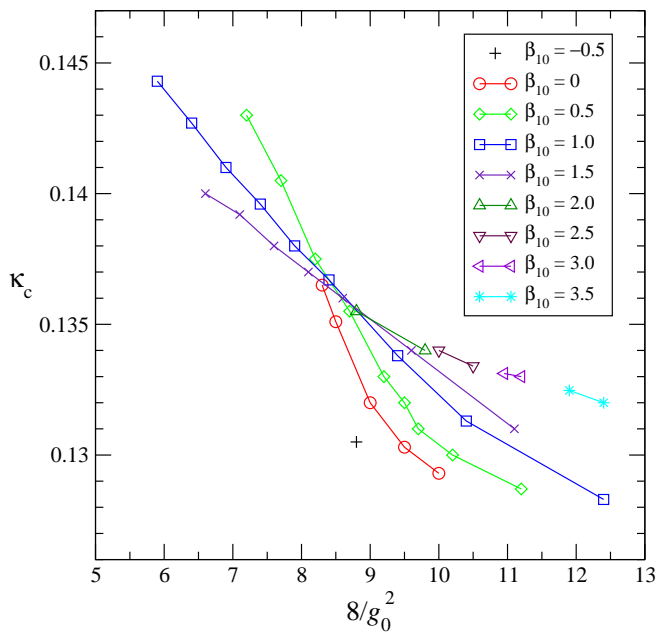}
\end{center}
\caption{Dependence of $\kappa_c$ on bare coupling
$8/g_0^2=\beta+\frac{12}{5}\beta_{10}$, for various values of $\beta_{10}$.
\label{fig:bkshift}}
\end{figure}

Finally, the simulations we are performing with the soft action are done at rather strong coupling.
In this regime, perturbation theory does not reliably predict the observed additive mass renormalization.
This is shown indirectly in Fig.~\ref{fig:bkshift}, a plot of $\kappa_c$ against the bare coupling $8/g_0^2=\beta+\frac{12}5\beta_{10}$.
Different plotting symbols show different values of $\beta_{10}$. In weak coupling the observed
additive mass renormalization qualitatively agrees with what is shown in Fig.~\ref{fig:s0}: larger
$\beta_{10}/\beta$ produces larger additive mass renormalization. However, in strong coupling the effect
reverses. One cannot help speculating that there is a connection between a large value of $\kappa_c$ and
a nearby first order transition, and that when $\kappa_c$ falls, the transition has moved away.

\section{Ensembles \label{sec:ensembles}}

\begin{table}
\caption{Ensembles generated at the bare couplings
  $(\beta, \beta_{10}, \kappa_c)$ for the lattice sizes $L$ used in this study.
  Listed are the total number of trajectories for all streams at given
$(\beta,\beta_{10})$ and~$L$,
  the trajectory length, and the HMC acceptance.}
\begin{center}
\begin{ruledtabular}
\begin{tabular}{dcdrrcc}
  \beta &
  $\beta_{10}$ &
  \kappa_c  &
  $L/a$  &
  trajectories&
  trajectory&
  acceptance\\
&&&&(thousands)&length&\\
\hline
10.0 & 0 & 0.1293  &  6 & 19.6 & 1.0 & 0.97 \\ 
  &      &         &  8 &  6.9 & 1.0  & 0.95 \\ 
  &      &         & 12 &  4.8 & 1.0  & 0.94\\[2pt] 

 9.0 & 0 & 0.13206 &  6 & 20.2 & 1.0  & 0.96 \\ 
  &      &         &  8 &  8.1 & 1.0  & 0.95\\ 
  &      &         & 12 & 12.0 & 1.0  & 0.85\\ 
  &      &         & 16 &  5.4 & 1.0  & 0.78\\[2pt] 

 8.5 & 0 & 0.1349  &  6 & 20.2 &  1.0 & 0.89  \\ 
  &      &         &  8 & 13.6 & 1.0  & 0.78\\ 
  &      &         & 12 &  6.1 & 0.75 & 0.66\\[2pt] 

 7.0 & 1 & 0.13361 &  6 & 48.8 & 1.0 & 0.95 \\ 
  &      &         &  8 & 16.4 & 1.0 & 0.93 \\ 
  &      &         & 12 & 10.3 & 1.0  & 0.68\\[2pt] 

 6.0 & 1 & 0.13634 &  6 & 10.0 & 1.0 & 0.82 \\ 
  &      &         &  8 & 20.0 & 0.5  & 0.88\\ 
  &      &         & 12 &  9.6 & 0.5  & 0.61\\[2pt] 

 5.0 & 1 & 0.13912 &  6 &  8.6 & 1.0 & 0.55 \\ 
  &      &         &  8 &  6.0 & 1.0 & 0.62 \\ 
  &      &         & 12 & 12.5 & 0.5  & 0.48\\ 
  &      &         & 16 & 10.7 & 0.5  & 0.63\\[2pt] 

 4.5 & 1 & 0.14055 &  6 &  8.6 & 1.0 & 0.52 \\ 
  &      &         &  8 & 13.0 & 0.5  & 0.84\\ 
  &      &         & 12 & 10.6 & 0.5  & 0.56\\ 
  &      &         & 16 &  9.7 & 0.5  & 0.45\\[2pt] 

 4.0 & 1 & 0.14193 &  6 &  9.4 & 1.0  & 0.58 \\ 
  &      &         &  8 & 12.0 & 0.5  & 0.72\\ 
  &      &         & 12 & 16.2 & 0.5  & 0.43\\ 
  &      &         & 16 & 20.2 & 0.4  & 0.37\\[2pt] 

 3.5 & 1 & 0.14318 &  6 & 14.6 & 1.0 & 0.53 \\ 
  &      &         &  8 &  9.4 & 0.5 & 0.66 \\ 
  &      &         & 12 & 19.2 & 0.25 & 0.64\\ 
\end{tabular}
\end{ruledtabular}
\end{center}
\label{table:runs}
\end{table}
Our algorithm is the hybrid Monte Carlo (HMC) algorithm of Duane and
Kogut~\cite{Duane:1985hz,Duane:1986iw,Gottlieb:1987mq}.
We accelerate the molecular dynamics integration with an additional heavy pseudo-fermion
field as suggested by Hasenbusch~\cite{Hasenbusch:2001ne}, with multiple time scales~\cite{Urbach:2005ji},
and with a second-order Omelyan integrator~\cite{Takaishi:2005tz}.
Lattice sizes range from $6^4$ to $16^4$ sites.

We list in Table~\ref{table:runs} the values of $(\beta,\beta_{10},\kappa_c)$ and the
number of trajectories run at each volume, along with the
length of the trajectories and the acceptance.
Poor acceptance forced us to shorten the trajectory length in many cases from the usual value of 1.

The observables we measure are the (inverse)
Schr\"odinger-functional running coupling, $1/g^2$,
and the pseudoscalar renormalization factor, $Z_P$.
(We measure $Z_P$ on the same configurations used to determine $1/g^2$.)
Both of them turn out to have long autocorrelations.
We monitored and controlled this problem by running 4 or 8 streams in parallel at each $(\beta,\beta_{10})$ and $L$.
After analyzing each stream separately,
we fit the results of
the streams together to a constant.
We demanded that the $\chi^2/{\rm dof}$
of the constant fit not exceed 6/3 for 4 streams, or 10/7 for 8 streams.
For the largest volume $L=16a$ at $(\beta,\beta_{10})=(5,1)$,
we were not able to overcome the autocorrelations in $1/g^2$.
We therefore omit this point from the analysis of the running coupling.
The autocorrelations in $Z_P$, on the other hand, did allow a consistent
determination, and thus
we keep this point in the analysis of the mass anomalous dimension.



\begin{thebibliography}{99}




\bibitem{reviews}
For reviews, see
  G.~T.~Fleming,
  PoS LATTICE {\bf 2008}, 021 (2008)
  [arXiv:0812.2035 [hep-lat]];
%
  E.~Pallante,
  PoS LAT {\bf 2009}, 015 (2009)
  [arXiv:0912.5188 [hep-lat]];
%
  L.~Del Debbio,
  PoS LATTICE {\bf 2010}, 004 (2010).

\bibitem{Hill:2002ap}
For a review of the phenomenology, see
  C.~T.~Hill and E.~H.~Simmons,
  ``Strong dynamics and electroweak symmetry breaking,''
  Phys.\ Rept.\  {\bf 381}, 235 (2003)
  [Erratum-ibid.\  {\bf 390}, 553 (2004)]
  [arXiv:hep-ph/0203079].


\bibitem{Caswell:1974gg}
  W.~E.~Caswell,
  ``Asymptotic behavior of nonabelian gauge theories to two loop order,''
  Phys.\ Rev.\ Lett.\  {\bf 33}, 244 (1974).

\bibitem{Banks:1981nn}
  T.~Banks and A.~Zaks,
  ``On the phase structure of vector-like gauge theories with massless fermions,''
  Nucl.\ Phys.\  B {\bf 196}, 189 (1982).


\bibitem{Shamir:2008pb}
  Y.~Shamir, B.~Svetitsky and T.~DeGrand,
  ``Zero of the discrete beta function in SU(3) lattice gauge theory with color sextet fermions,''
  Phys.\ Rev.\  D {\bf 78}, 031502 (2008)
  [arXiv:0803.1707 [hep-lat]].

\bibitem{DeGrand:2008kx}
  T.~DeGrand, Y.~Shamir and B.~Svetitsky,
  ``Phase structure of SU(3) gauge theory with two flavors of symmetric-representation fermions,''
  Phys.\ Rev.\  D {\bf 79}, 034501 (2009)
  [arXiv:0812.1427 [hep-lat]].

\bibitem{DeGrand:2009hu}
  T.~DeGrand,
  ``Finite-size scaling tests for SU(3) lattice gauge theory with color sextet fermions,''
  Phys.\ Rev.\  D {\bf 80}, 114507 (2009)
  [arXiv:0910.3072 [hep-lat]].

\bibitem{DeGrand:2010na}
  T.~DeGrand, Y.~Shamir, B.~Svetitsky,
  ``Running coupling and mass anomalous dimension of SU(3) gauge theory with two flavors of symmetric-representation fermions,''
  Phys.\ Rev.\  {\bf D82}, 054503 (2010)
  [arXiv:1006.0707 [hep-lat]].

\bibitem{DeGrand:2011qd}
  T.~DeGrand, Y.~Shamir, B.~Svetitsky,
  ``Infrared fixed point in SU(2) gauge theory with adjoint fermions,''
  Phys.\ Rev.\  {\bf D83}, 074507 (2011)
  [arXiv:1102.2843 [hep-lat]].

\bibitem{DeGrand:2011vp}
  T.~DeGrand, Y.~Shamir, B.~Svetitsky,
  ``Gauge theories with fermions in the two-index symmetric representation,''
  PoS LATTICE {\bf 2011}, 060 (2011)
  [arXiv:1110.6845 [hep-lat]].

\bibitem{DeGrand:2012yq}
  T.~DeGrand, Y.~Shamir and B.~Svetitsky,
  ``Mass anomalous dimension in sextet QCD,''
  arXiv:1201.0935 [hep-lat].


\bibitem{Sannino:2004qp}
  F.~Sannino and K.~Tuominen,
  ``Orientifold theory dynamics and symmetry breaking,''
  Phys.\ Rev.\  D {\bf 71}, 051901 (2005)
  [arXiv:hep-ph/0405209].

\bibitem{Dietrich:2006cm}
  D.~D.~Dietrich and F.~Sannino,
  ``Conformal window of SU($N$) gauge theories with fermions in higher dimensional representations,''
  Phys.\ Rev.\  D {\bf 75}, 085018 (2007)
  [arXiv:hep-ph/0611341].


\bibitem{Luscher:1992an}
  M.~L\"uscher, R.~Narayanan, P.~Weisz and U.~Wolff,
  ``The Schrodinger functional: A Renormalizable probe for nonAbelian gauge
  theories,''
  Nucl.\ Phys.\  B {\bf 384}, 168 (1992)
  [arXiv:hep-lat/9207009].

\bibitem{Luscher:1993gh}
  M.~L\"uscher, R.~Sommer, P.~Weisz and U.~Wolff,
  ``A precise determination of the running coupling in the SU(3) Yang-Mills theory,''
  Nucl.\ Phys.\  B {\bf 413}, 481 (1994)
  [arXiv:hep-lat/9309005].

\bibitem{Sint:1995ch}
  S.~Sint and R.~Sommer,
  ``The running coupling from the QCD Schr\"odinger functional: A one loop analysis,''
  Nucl.\ Phys.\  B {\bf 465}, 71 (1996)
  [arXiv:hep-lat/9508012].

\bibitem{Jansen:1998mx}
  K.~Jansen and R.~Sommer  [ALPHA collaboration],
  ``O($\alpha$) improvement of lattice QCD with two flavors of Wilson quarks,''
  Nucl.\ Phys.\  B {\bf 530}, 185 (1998)
  [Erratum-{\em ibid.}\  B {\bf 643}, 517 (2002)]
  [arXiv:hep-lat/9803017].

\bibitem{DellaMorte:2004bc}
  M.~Della Morte {\em et al.} [ALPHA Collaboration],
  ``Computation of the strong coupling in QCD with two dynamical flavours,''
  Nucl.\ Phys.\  B {\bf 713}, 378 (2005)
  [arXiv:hep-lat/0411025].


\bibitem{Sint:1998iq}
  S.~Sint and P.~Weisz  [ALPHA collaboration],
  ``The running quark mass in the SF scheme and its two-loop anomalous dimension,''
  Nucl.\ Phys.\  B {\bf 545}, 529 (1999)
  [arXiv:hep-lat/9808013].

\bibitem{Capitani:1998mq}
  S.~Capitani, M.~L\"uscher, R.~Sommer and H.~Wittig  [ALPHA Collaboration],
  ``Non-perturbative quark mass renormalization in quenched lattice QCD,''
  Nucl.\ Phys.\  B {\bf 544}, 669 (1999)
  [arXiv:hep-lat/9810063].

\bibitem{DellaMorte:2005kg}
  M.~Della Morte {\em et al.} 
  [ALPHA Collaboration],
  ``Non-perturbative quark mass renormalization in two-flavor QCD,''
  Nucl.\ Phys.\  B {\bf 729}, 117 (2005)
  [arXiv:hep-lat/0507035].


\bibitem{Lucini:2008vi}
  B.~Lucini and G.~Moraitis,
  ``The running of the coupling in SU($N$) pure gauge theories,''
  Phys.\ Lett.\ B {\bf 668}, 226 (2008)
  [arXiv:0805.2913 [hep-lat]].


\bibitem{Appelquist:2007hu}
  T.~Appelquist, G.~T.~Fleming and E.~T.~Neil,
  \ttl{Lattice study of the conformal window in QCD-like theories,}
  Phys.\ Rev.\ Lett.\  {\bf 100}, 171607 (2008)
  [Erratum-ibid.\  {\bf 102}, 149902 (2009)]
  [arXiv:0712.0609 [hep-ph]].

\bibitem{Appelquist:2009ty}
  T.~Appelquist, G.~T.~Fleming and E.~T.~Neil,
  \ttl{Lattice Study of Conformal Behavior in SU(3) Yang-Mills Theories,}
  Phys.\ Rev.\ D {\bf 79}, 076010 (2009)
  [arXiv:0901.3766 [hep-ph]].

\bibitem{Hietanen:2009az}
  A.~J.~Hietanen, K.~Rummukainen and K.~Tuominen,
  \ttl{Evolution of the coupling constant in SU(2) lattice gauge theory with two adjoint fermions,}
  Phys.\ Rev.\ D {\bf 80}, 094504 (2009)
  [arXiv:0904.0864 [hep-lat]].

\bibitem{Bursa:2009we}
  F.~Bursa, L.~Del Debbio, L.~Keegan, C.~Pica and T.~Pickup,
  \ttl{Mass anomalous dimension in SU(2) with two adjoint fermions,}
  Phys.\ Rev.\  D {\bf 81}, 014505 (2010)
  [arXiv:0910.4535 [hep-ph]].

\bibitem{Bursa:2010xn}
  F.~Bursa, L.~Del Debbio, L.~Keegan, C.~Pica and T.~Pickup,
  \ttl{Mass anomalous dimension in SU(2) with six fundamental fermions,}
  Phys.\ Lett.\ B {\bf 696}, 374 (2011)
  [arXiv:1007.3067 [hep-ph]].

\bibitem{Hayakawa:2010yn}
  M.~Hayakawa, K.~-I.~Ishikawa, Y.~Osaki, S.~Takeda, S.~Uno and N.~Yamada,
  \ttl{Running coupling constant of ten-flavor QCD with the Schr\"odinger functional method,}
  Phys.\ Rev.\ D {\bf 83}, 074509 (2011)
  [arXiv:1011.2577 [hep-lat]].

\bibitem{Karavirta:2011zg}
  T.~Karavirta, J.~Rantaharju, K.~Rummukainen and K.~Tuominen,
  \ttl{Determining the conformal window: SU(2) gauge theory with $N_f = 4$, 6 and 10 fermion flavours,}
  arXiv:1111.4104 [hep-lat];
%
  ``Exploring the conformal window: SU(2) gauge theory on the lattice,''
  arXiv:1201.2037 [hep-lat].

\bibitem{Karavirta:2012qd}
  T.~Karavirta, K.~Rummukainen and K.~Tuominen,
  ``Perturbative Improvement of the Schrodinger Functional for Lattice Strong Dynamics,''
  arXiv:1201.1883 [hep-lat].

\bibitem{DeGrand:2009mt}
  T.~DeGrand and A.~Hasenfratz,
  ``Remarks on lattice gauge theories with infrared-attractive fixed points,''
  Phys.\ Rev.\ D {\bf 80}, 034506 (2009)
  [arXiv:0906.1976 [hep-lat]].



\bibitem{Sheikholeslami:1985ij}
  B.~Sheikholeslami and R.~Wohlert,
  ``Improved continuum limit lattice action for QCD with Wilson fermions,''
  Nucl.\ Phys.\  B {\bf 259}, 572 (1985).

\bibitem{Hasenfratz:2001hp}
  A.~Hasenfratz and F.~Knechtli,
  ``Flavor symmetry and the static potential with hypercubic blocking,''
  Phys.\ Rev.\  D {\bf 64}, 034504 (2001)
  [arXiv:hep-lat/0103029].

\bibitem{Hasenfratz:2007rf}
  A.~Hasenfratz, R.~Hoffmann and S.~Schaefer,
  ``Hypercubic smeared links for dynamical fermions,''
  JHEP {\bf 0705}, 029 (2007)
  [arXiv:hep-lat/0702028].

\bibitem{Iwasaki:1991mr}
  Y.~Iwasaki, K.~Kanaya, S.~Sakai and T.~Yoshi\'e,
  ``Quark confinement and number of flavors in strong coupling lattice QCD,''
  Phys.\ Rev.\ Lett.\  {\bf 69}, 21 (1992).

\bibitem{Iwasaki:2003de}
  Y.~Iwasaki, K.~Kanaya, S.~Kaya, S.~Sakai and T.~Yoshi\'e,
  ``Phase structure of lattice QCD for general number of flavors,''
  Phys.\ Rev.\  D {\bf 69}, 014507 (2004)
  [arXiv:hep-lat/0309159].

\bibitem{Nagai:2009ip}
  K.~Nagai, G.~Carrillo-Ruiz, G.~Koleva and R.~Lewis,
  ``Exploration of SU($N_c$) gauge theory with many Wilson fermions at strong coupling,''
  Phys.\ Rev.\  D {\bf 80}, 074508 (2009)
  [arXiv:0908.0166 [hep-lat]].

\bibitem{Catterall:2008qk}
  S.~Catterall, J.~Giedt, F.~Sannino and J.~Schneible,
  ``Phase diagram of SU(2) with 2 flavors of dynamical adjoint quarks,''
  JHEP {\bf 0811}, 009 (2008)
  [arXiv:0807.0792 [hep-lat]].

\bibitem{Hietanen:2008mr} 
  A.~J.~Hietanen, J.~Rantaharju, K.~Rummukainen and K.~Tuominen,
  JHEP {\bf 0905}, 025 (2009)
  [arXiv:0812.1467 [hep-lat]].
\bibitem{MILC} {\tt http://www.physics.utah.edu/$\sim$detar/milc/}

\bibitem{Hasenfratz:2008ce}
  A.~Hasenfratz, R.~Hoffmann and S.~Schaefer,
  ``Low energy chiral constants from epsilon-regime simulations with improved Wilson fermions,''
  Phys.\ Rev.\  D {\bf 78}, 054511 (2008)
  [arXiv:0806.4586 [hep-lat]].

\bibitem{Morningstar:2003gk}
  C.~Morningstar and M.~J.~Peardon,
  ``Analytic smearing of SU(3) link variables in lattice QCD,''
  Phys.\ Rev.\ D {\bf 69}, 054501 (2004)
  [hep-lat/0311018].

\bibitem{ancillary}
See the ancillary material accompanying this paper in the archive for
a {\em Mathematica\/} notebook for calculating $f_i$ and $b_{ij}$, as well as a PDF file of the results.

\bibitem{Bernard:1999kc}
  C.~W.~Bernard and T.~A.~DeGrand,
  ``Perturbation theory for fat link fermion actions,''
  Nucl.\ Phys.\ Proc.\ Suppl.\  {\bf 83}, 845 (2000)
  [arXiv:hep-lat/9909083].
  See also
  T.~A.~DeGrand,
  ``One loop matching coefficients for a variant overlap action and some of its simpler relatives,''
  Phys.\ Rev.\  D {\bf 67}, 014507 (2003)
  [arXiv:hep-lat/0210028].

\bibitem{Hasenfratz:2001tw}
  A.~Hasenfratz, R.~Hoffmann and F.~Knechtli,
  ``The Static potential with hypercubic blocking,''
  Nucl.\ Phys.\ Proc.\ Suppl.\  {\bf 106}, 418 (2002)
  [hep-lat/0110168].

\bibitem{Duane:1985hz}
  S.~Duane and J.~B.~Kogut,
  ``Hybrid Stochastic Differential Equations Applied to Quantum Chromodynamics,''
  Phys.\ Rev.\ Lett.\  {\bf 55}, 2774 (1985).

\bibitem{Duane:1986iw}
  S.~Duane and J.~B.~Kogut,
  ``The Theory Of Hybrid Stochastic Algorithms,''
  Nucl.\ Phys.\ B {\bf 275}, 398 (1986).

\bibitem{Gottlieb:1987mq}
  S.~A.~Gottlieb, W.~Liu, D.~Toussaint, R.~L.~Renken and R.~L.~Sugar,
  ``Hybrid Molecular Dynamics Algorithms for the Numerical Simulation of Quantum Chromodynamics,''
  Phys.\ Rev.\ D {\bf 35}, 2531 (1987).

\bibitem{Hasenbusch:2001ne}
  M.~Hasenbusch,
  ``Speeding up the Hybrid-Monte-Carlo algorithm for dynamical fermions,''
  Phys.\ Lett.\  B {\bf 519}, 177 (2001)
  [arXiv:hep-lat/0107019].

\bibitem{Urbach:2005ji}
 C.~Urbach, K.~Jansen, A.~Shindler and U.~Wenger,
 ``HMC algorithm with multiple time scale integration and mass preconditioning,''
 Comput.\ Phys.\ Commun.\  {\bf 174}, 87 (2006)
 [arXiv:hep-lat/0506011].

\bibitem{Takaishi:2005tz}
  T.~Takaishi and P.~de Forcrand,
  ``Testing and tuning new symplectic integrators for hybrid Monte Carlo algorithm in lattice QCD,''
  Phys.\ Rev.\  E {\bf 73}, 036706 (2006)
  [arXiv:hep-lat/0505020].
  

\end{thebibliography}
\end{document}